\documentclass[12pt]{article}
\pdfoutput=1
\usepackage{amsmath,amsfonts,graphicx,color,bbm,tikz,bm,setspace}
\usepackage{comment}
\usepackage[nosort]{cite}
\usepackage{subfigure}
\usepackage{multirow}
\usetikzlibrary{calc,positioning}
\usetikzlibrary{patterns,arrows,decorations.pathreplacing}
\usepackage{caption}
\usepackage{ulem}
\tikzset{>=stealth}
\usepackage{lipsum}
\usepackage{tabularx}
\usepackage{longtable}
\usepackage{fullpage}
\usepackage{hyperref}
\usepackage{bbold}
\usepackage{arydshln}
\usepackage[english]{babel}
\newtheorem{definition}{Definition}[section]

\newtheorem{remark}{Remark}[section]

\usepackage{listings}
\usepackage{tcolorbox}

\newcommand{\mathsym}[1]{{}}
\newcommand{\unicode}[1]{{}}

\makeatletter\@addtoreset{equation}{section}\makeatother

\newcommand{\be}{\begin{equation}}
\newcommand{\ee}{\end{equation}}
\def\beq{\begin{equation}}
\def\eeq{\end{equation}}
\newcommand{\bea}{\begin{eqnarray}}
\newcommand{\eea}{\end{eqnarray}}

\newcommand{\tr}{{\rm tr\,}}

\newcommand{\ket}[1]{{\left| {#1} \right>}}

\renewcommand{\title}[1]{\vbox{\center\LARGE{#1}}\vspace{3mm}}
\renewcommand{\author}[1]{\vbox{\center{#1}}\vspace{3mm}}

\newcommand{\email}[1]{\vbox{\center\tt#1}\vspace{3mm}}

\begin{document}
%\begin{titlepage}

\begin{center}
{\large {\bf Almost local integrable models from supersymmetry algebras}}

\author{Somnath Maity,$^a$ Pramod Padmanabhan,$^a$ Jarmo Hietarinta,$^{b}$ Vladimir Korepin$^{c}$}

\vskip 0.05cm
{$^a${\it Department of Physics, School of Basic Sciences,\\ Indian Institute of Technology, Bhubaneswar, 752050, India}}
\vskip0.1cm
{$^b${\it Department of Physics and Astronomy, \\ University of Turku, \\ FIN-20014 Turku, Finland}}
\vskip0.1cm
{$^c${\it C. N. Yang Institute for Theoretical Physics, \\ Stony Brook University, New York 11794, USA}}

\email{somnathmaity126@gmail.com, hietarin@utu.fi, vladimir.korepin@stonybrook.edu}

 \begin{center}
\textbf{Corresponding author:} Pramod Padmanabhan (\texttt{pramod23phys@gmail.com})
\end{center}

\today
%\vskip 0.2cm 

\end{center}

%%%%%%%%%%%

\abstract{
\noindent 
Supersymmetry algebras can be used to obtain algebraic expressions for constant Yang-Baxter solutions, also known as braid group generators. This was done for non-invertible braid operators in \cite{maity2025non}. In this work we extend this construction for the invertible ones. The resulting expressions are then shown to obey relations analogous to those satisfied by quotients of braid groups. Examples of the latter include the Iwahori-Hecke algebra and the Birman-Murakami-Wenzl (BMW) algebra. As a result, we can Baxterize the constant Yang-Baxter solutions to yield spectral parameter dependent $R$-matrices. The regularity of these $R$-matrices depends on the representation of SUSY generators. In some cases they are regular in the usual sense and in the remaining they are `almost' regular. In the latter case they are also non-invertible. Nevertheless, we show that they can still help us construct integrable models in all dimensions of the local Hilbert space. These models can be described by Hamiltonian densities that are either local or non-local, depending on the representation chosen for the SUSY generators. We demonstrate this for all constant $4\times 4$ invertible Yang-Baxter solutions. Apart from finding new nearest-neighbor interaction spin $\frac{1}{2}$ systems, we also find their higher spin analogs due to the algebraic [representation independent] approach.
}

%\end{titlepage}
\tableofcontents 

%%%%%%%%%%%%%%%%%%%%%%%%%%
\section{Introduction}
\label{sec:Introduction}
%%%%%%%%%%%%%%%%%%%%%%%%%%
Among the many notions of quantum integrability \cite{Caux_2011} a well explored one is the integrability arising out of the quantum Yang-Baxter equation \cite{BAXTER1972193,YangCN1967}. This approach systematically leads to the construction of many mutually commuting quantities \cite{loebbert2016lectures} that have also shown to be useful in the simulations of these models on a quantum computer \cite{Vanicat2017IntegrableTL}. One approach to construct such integrable models is to Baxterize constant quantum Yang-Baxter solutions, also known as braid operators\footnote{As they generate the braid group.}\footnote{Algebras, derived from the braid group algebras can be Baxterized as well \cite{Crampe_2016,crampe_2019}.} \cite{Jones1990,li1993yang,ge-baxterization,kulish-baxterization,ZHANG1991625}. A great opportunity for this is presented in the early works of Jarmo Hietarinta, where he exhausted all possible [both invertible and non-invertible] $4\times 4$ constant quantum Yang-Baxter solutions \cite{HIETARINTA-PLA,hietarinta1993-JMP-Long}, after several earlier attempts to do the same \cite{Hlavaty_1987,Fei:1991zn,Sogo,Zhang_1991}. Following this pioneering work, there were almost no successful effort to Baxterize these solutions systematically. In this work we revive this program albeit using an algebraic approach, that is in a representation independent manner. For this we resort to using $\mathcal{N}=2$ supersymmetry (SUSY) algebras to systematically construct the algebraic expressions for the quantum Yang-Baxter solutions. The resulting expressions are invertible or non-invertible depending on the representation chosen for the SUSY generators. Furthermore, the algebraic expressions for these constant operators are shown to obey additional constraints implying that they do not generate the infinite dimensional braid (semi)group. Instead they realize quotients much like the Iwahori-Hecke algebra. In this manner we generalize the results obtained in our previous work \cite{maity2025non}, where we Baxterize the $4\times 4$ constant non-invertible solutions found in \cite{HIETARINTA-PLA}. As a result of this process we find $R$-matrices that are either invertible or non-invertible depending on the representation chosen for the SUSY algebra. The resulting spin chains are a sum of local [nearest-neighbor] or global [support on $L$ sites of the chain]. For this reason we dub them as {\it `almost local'}\footnote{Integrable models with medium to long range interactions can be found in 
\cite{bargheer2009long,Pozsgay:2021rwc}. They also appear in the context of the $AdS/CFT$ correspondence \cite{Kruczenski2004,minahan2012review,staudacher2012review}.}. Nevertheless the non-invertibility and the subsequent non-locality appear in a controlled manner where many of the usual techniques used in the quantum inverse scattering method \cite{takhtadzhyan1979,Takhtadzhan_1979,Korepin1993QuantumIS} still go through with very few hiccups. Thus apart from systematically constructing spin chains in all dimensions, we also develop theoretical tools that enhance the scope of the quantum Yang-Baxter equation which we hope will spur new studies in this field. We will now go over the basics of the quantum Yang-Baxter equation before laying out the contents of this paper.

%%%%%%%%%%%%%%%%%%%%%%%%%%%%%%%%%%%%%%%%%%
\subsection{Quantum Yang-Baxter equation}
\label{subsec:QYBE}
%%%%%%%%%%%%%%%%%%%%%%%%%%%%%%%%%%%%%%%%%%%
The quantum Yang-Baxter equation (QYBE) appears in two forms according to the index structure on the operators. One of them is the so called non-braided form of the QYBE
\begin{equation}\label{eq:non-braided-YBE}
    R_{12}(u)R_{13}(u+v)R_{23}(v) = R_{23}(v)R_{13}(u+v)R_{12}(u),
\end{equation}
is most suited for the construction of quantum integrable systems.
This is an operator-valued equation that acts on $V \otimes V \otimes V$ with $V$ being a local Hilbert space. The operator $R_{ij}(u)\in \mathrm{End}(V\otimes V)$ acts non-trivially on the $i^{th}$ and  $j^{th}$ components of the full tensor product space. The above form is also referred to as the additive form of the QYBE due to the particular form in which the complex spectral parameters $u$, $v$ appear. There is also the non-additive form, 
$   R_{12}(u_1,u_2)R_{13}(u_1,u_3)R_{23}(u_2,u_3) = R_{23}(u_2,u_3)R_{13}(u_1,u_3)R_{12}(u_1,u_2),$
but we will not use it here. It is applicable, for instance in the chiral Potts model \cite{BAXTER1988138,AuYang2016About3Y}.

The second type, which will be the focus of this work, is called the braided form of the QYBE
\begin{equation}\label{eq:braided-YBE}
    \check{R}_{12}(u)\check{R}_{23}(u+v)\check{R}_{12}(v) = \check{R}_{23}(v)\check{R}_{12}(u+v)\check{R}_{23}(u).
\end{equation}
The above types of QYBE are related: The braided form can be obtained by multiplying the permutation operator $P_{ij}$ with $R_{ij}$,
\begin{eqnarray}
    \check{R}_{ij} = P_{ij}\cdot R_{ij}.
\end{eqnarray}
Using the notation $P_i\equiv P_{i,i+1}$, the permutation operator $P:V_i \otimes V_j \to V_j \otimes V_i$ satisfies the relations
\begin{eqnarray}\label{eq:perm-relations}
    P_iP_{i+1}P_i = P_{i+1}P_iP_{i+1}~;~P^2=\mathbb{1}~;~P_iP_j = P_jP_i,~~|i-j|>1,
\end{eqnarray}
of the generators of the permutation group. 

For certain special values of the spectral parameters the above QYBE's reduce to the so called constant YBE's, also known as the {\it braid relations} due to their relation to the braid group. These $R$-matrices are called constant solutions and they obey the equation 
\begin{equation}\label{eq:braid-relation}
    \check{Y}_{12} \check{Y}_{23} \check{Y}_{12} = \check{Y}_{23} \check{Y}_{12} \check{Y}_{23}.
\end{equation}
Multiplying these $\check{Y}$ operators by the permutation operator $P$ results in the non-braided form of the constant QYBE.

%%%%%%%%%%%%%%%%%%%%%%%%%%%%%%%%%%%%
\subsection{Organization}
\label{subsec:plan}
%%%%%%%%%%%%%%%%%%%%%%%%%%%%%%%%%%%%
The basic ingredients needed to carry out the construction of the integrable models is detailed in a preliminary section \ref{sec:preliminaries}. This is followed by the various Baxterization procedures in Sec.\ \ref{sec:Baxterization}. They facilitate the construction of the integrable models presented in this paper. The methods developed are then applied on the $4\times 4$ constant solutions of Hietarinta, split into two cases : diagonalizable [Sec.\ \ref{sec:Models-diagonalizable}] and non-diagonalizable [Sec.\ \ref{sec:Models-nondiagonalizable}], as the Baxterization methods for the two are different. The resulting Hamiltonians and some of their properties are summarized in Sec.\ \ref{sec:summary}. This section also includes the new spin $\frac{1}{2}$ nearest-neighbor spin chains as shown in Table \ref{tab:new-h}.  An important feature of our construction is that it is algebraic. This implies that we get a new solution in each dimension of the local Hilbert space by choosing an appropriate representation of the SUSY algebra generators. In the higher spin case we obtain several inequivalent spin chain Hamiltonians whose number coincides with the number of inequivalent representations of the generators of the SUSY algebra. This construction is shown in Sec.\ \ref{sec:higher-spin}. Some final remarks are added in a short conclusion found in Sec.\ \ref{sec:conclusion}. 

%A short appendix defining the standard Iwahori-Hecke algebra is in App. \ref{app:Hecke-algebra}. The modified boost operator method for the `almost local' Hamiltonians derived in this work is explained in detail in App. \ref{app:boost}.

%%%%%%%%%%%%%%%%%%%%%%%%%%%%%%%%%%%%%%%%%%%%%%%%%%%%%%%%%%%%%%%%%%%%%%
\section{Preliminaries}
\label{sec:preliminaries}
%%%%%%%%%%%%%%%%%%%%%%%%%%%%%%%%%%%%%%%%%%%%%%%%%%%%%%%%%%%%%%%%%%%%%%
The goal of this work is to construct spectral parameter dependent Yang-Baxter solutions, or $R$-matrices, in a representation independent manner using supersymmetry (SUSY) algebras. We then want to compare the integrable models obtained from these $R$-matrices with known models in the literature and identify the new models in our set of solutions. To achieve this we need to understand the following three ingredients :
\begin{enumerate}
    \item $\mathcal{N}=2$ SUSY algebras and how they realize the algebra of $2\times 2$ matrices ${\mathcal Mat}(2,\mathbb{C})$.
    \item Derivation of integrable Hamiltonians from $R$-matrices constructed out of such SUSY algebras.
    \item Definitions of equivalence classes of $R$-matrices and equivalence classes of local Hamiltonian densities.
\end{enumerate}

%%%%%%%%%%%%%%%%%%%%%%%%%%%%%%%%%%%%%%%%%%%%%%%%%%%%%%%%%%%%%%%%%%%%%%
\subsection{$\mathcal{N}=2$ supersymmetry (SUSY) algebra}
\label{subsec:susy-algebra}
%%%%%%%%%%%%%%%%%%%%%%%%%%%%%%%%%%%%%%%%%%%%%%%%%%%%%%%%%%%%%%%%%%%%%%
The $\mathcal{N}=2$ SUSY algebra is generated by the supercharges $q$ and $q^\dag$ [See {\it supersymmetric quantum mechanics} \cite{Witten1981DynamicalBO,Cooper_1995}], obeying the relations
\begin{eqnarray}\label{eq:susy-relations}
    &q^2=\left(q^\dag\right)^2=0,&\\
    &h:=\left\{q, q^\dag\right\}.&
\end{eqnarray}
Here $h$ is called the SUSY Hamiltonian\footnote{These Hamiltonians are different from the spin chain Hamiltonians $H$, constructed later.} as it commutes with the supercharges
\begin{eqnarray}
    \left[q, h\right] = \left[q^\dag, h\right] = 0.
\end{eqnarray}
The representations of this algebra that are useful for our purposes regard these relations as realizing a $\mathbb{Z}_2$ grading of a $d$-dimensional Hilbert space $\mathbb{C}^d$. In this context the operators 
\begin{equation*}
    b: = qq^\dag,\quad f:=q^\dag q,
\end{equation*}
are seen as projectors to the two parts of the graded Hilbert space. Traditionally, these two parts are called the `Bosonic' and `Fermionic' sectors respectively. For us these terms are not relevant in their intended sense but nevertheless we will use them for convenience. This also explains the notation we have chosen for the two projectors respectively. The supercharges are operators transforming between the two sectors. A sufficient condition for operators $b$ and $f$ to be projectors is 
\begin{equation}\label{eq:qqdagq}
    qq^\dag q=q,\quad q^\dag q q^\dag =q^\dag.
\end{equation}
Note that these relations need not hold for a general SUSY algebra as they do not follow from the SUSY relations in \eqref{eq:susy-relations}.
As a consequence, we see that the SUSY Hamiltonian is a projector $$h^2=h,$$ in an arbitrary $d$-dimensional representation. There are two possible choices for $h$ in a given representation. It is either a $d$-dimensional identity operator $\mathbb{1}_{d\times d}$ or a non-trivial projector with at least one of its eigenvalues as zero. We will encounter both cases in the representations we use. In both these possibilities the following relations hold
\begin{equation}\label{eq:susy-h-identity-like}
     hq=qh=q,\quad hq^\dag = q^\dag h = q^\dag,\quad hb=bh=b,\quad hf=fh=f.
\end{equation}

Thus in all the algebraic computations involving the elements of the SUSY algebra the SUSY Hamiltonian plays the role of exactly the identity operator or an `almost' identity operator. This aspect will be crucial in defining the `almost' regular $R$-matrix that we will soon encounter.

The 2-dimensional representation of the SUSY algebra in $\mathbb{C}^2$ helps establish an isomorphism between the matrix algebra ${\mathcal Mat}(2,\mathbb{C})$ and the SUSY algebra elements in \eqref{eq:susy-relations}. This is immediate from the explicit expression for the representation of the SUSY generators
\begin{equation}\label{eq:susy-representation-C2}
    q= \begin{pmatrix}
        0 & 1 \\ 0 & 0
    \end{pmatrix},\quad q^\dag = \begin{pmatrix}
        0 & 0 \\ 1 & 0
    \end{pmatrix},    \quad
      b= \begin{pmatrix}
        1 & 0 \\ 0 & 0
    \end{pmatrix},\quad f= \begin{pmatrix}
        0 & 0 \\ 0 & 1
    \end{pmatrix}.  
\end{equation}
The isomorphism is evident from the identification
\begin{eqnarray}
    q=e_{12},\quad q^\dag=e_{21},\quad b=e_{11},\quad f=e_{22},
\end{eqnarray}
where $e_{ij}=\delta_{ij}$ form the basis of the vector space underlying the matrix algebra ${\mathcal Mat}(2,\mathbb{C})$. Note also that using Pauli matrices,
\begin{eqnarray}\label{eq:q2sigma}
    q=\frac{\sigma^x+\mathrm{i}~\sigma^y}{2}=\sigma^+~;~q^\dag=\frac{\sigma^x-\mathrm{i}~\sigma^y}{2}=\sigma^-~;~b=\frac{\mathbb{1}+\sigma^z}{2}~;~f=\frac{\mathbb{1}-\sigma^z}{2}.
\end{eqnarray}
Thus any $2\times 2$ matrix can be cast in terms of the SUSY generators. This then extends to operators acting on tensor products of $\mathbb{C}^2$ and so this isomorphism will help us write Yang-Baxter solutions in terms of tensor products of
\begin{eqnarray}
  \left\{q, q^\dag, b, f\right\}.  
\end{eqnarray}
This property will serve as a starting point of converting the $4\times 4$ constant Yang-Baxter solutions into algebraic expressions in Sec.\ \ref{sec:Models-diagonalizable}. The resulting expressions for the Yang-Baxter solutions are guaranteed to be algebraic due to this isomorphism. So we can obtain the higher dimensional versions of these operators by inserting the corresponding higher dimensional representations of these SUSY generators. It should be noted that the representation of the SUSY generators in \eqref{eq:susy-representation-C2} satisfy the additional requirement on the supercharges \eqref{eq:qqdagq}. As a result the SUSY Hamiltonian is a projector, in fact the $2\times 2$ identity operator, as desired. 

\begin{remark}\label{rem:nilpotent-qs-c2}
    Any nilpotent $2\times 2$ matrix is admissible as a supercharge as it satisfies the SUSY relations in \eqref{eq:susy-relations}. However not all of them satisfy the additional constraint on the supercharges \eqref{eq:qqdagq}. We find that for $\alpha, \beta\in\mathbb{R}$,
    $$ q=\frac{1}{2\sqrt{\alpha^2+\beta^2}} \left[\pm\mathrm{i}\sqrt{\alpha^2+\beta^2}~\sigma^x + \alpha~\sigma^y + \beta~\sigma^z\right],$$
    satisfies the constraint \eqref{eq:qqdagq} and hence produces projector SUSY Hamiltonians that are just the $2\times 2$ identity operator. However these supercharges are equivalent to the supercharge in \eqref{eq:susy-representation-C2} by a similarity transform and so we do not consider them in constructing Yang-Baxter solutions and their associated integrable models.
\end{remark}

%%%%%%%%%%%%%%%%%%%%%%%%%%%%%%%%%%%%%%%%%%%%%%%%%%%%%%%%%%%%%%%%%%%%%%
\subsection{Integrable models from non-invertible $R$-matrices}
\label{subsec:integrable-noninvertible-R}
%%%%%%%%%%%%%%%%%%%%%%%%%%%%%%%%%%%%%%%%%%%%%%%%%%%%%%%%%%%%%%%%%%%%%%
%In this subsection we will use $\check{R}$, the non-braided $R$-matrix in place of the $R$ used so far. 
We see that due to the appearance of projector SUSY Hamiltonians $h$, the spectral parameter dependent Yang-Baxter solutions or the $R$-matrices are in general non-invertible. Nevertheless we can define an altered inverse that satisfies the relation
\begin{eqnarray}\label{eq:R-inverse-altered}
    R_{jk}(u)R^{-\iota}_{jk}(u)= R_{jk}^{-\iota}(u)R_{jk}(u) = h_jh_k~;~u\in\mathbb{C},
\end{eqnarray}
with $u$ the complex spectral parameter. Here the symbol $R^{-\iota}$ denotes the altered inverse. We also assume that the left and right inverses are the same. Also note that the altered inverse reduces to the usual inverse in the cases where the SUSY Hamiltonian is the identity operator $\mathbb{1}_{d\times d}$. This depends on the representation we choose for the supercharges. We will analyze the higher spin representations in Sec.\ \ref{sec:higher-spin}. 
In this situation an immediate question arises. 

{\it Does the monodromy matrix, constructed using such $R$-matrices and satisfying the $RTT$ relation, give rise to a set of mutually commuting operators on the physical Hilbert space?}

We will now see that this is indeed the case for the SUSY $R$-matrices with their altered inverses satisfying \eqref{eq:R-inverse-altered}. Consider the the monodromy matrix $T$ defined as
\begin{eqnarray}
    T_{j}(u):=R_{jL}(u)\cdots R_{j1}(u),
\end{eqnarray}
on a chain of $L$ sites. We require this operator to satisfy the $RTT$ relation :
\begin{eqnarray}\label{eq:RTT}
    R_{jk}(u-v)T_j(u)T_k(v) = T_k(v)T_j(u)R_{jk}(u-v),
\end{eqnarray}
Here $j$ and $k$ index auxiliary spaces \cite{slavnov2019algebraicbetheansatz,Korepin1993QuantumIS}, as opposed to the physical spaces indexed by the set $\left\{1,2\cdots,L \right\}$. The transfer matrix is obtained from the monodromy matrix by tracing out its auxiliary index,
\begin{equation}
    \mathcal{T}(u) := tr_j\left[T_j(u)\right] = tr_j[R_{jL}(u) \dots R_{jn}(u) \dots R_{j1}(u)].
\end{equation}
We will now show that this matrix commutes with itself evaluated at different values of the spectral parameter. To see this multiply both sides of the $RTT$ relation \eqref{eq:RTT} by $R^{-\iota}_{jk}(u-v)$, 
\begin{eqnarray}
     R_{jk}(u-v)T_j(u)T_k(v)R^{-\iota}_{jk}(u-v) & = & T_k(v)h_kT_j(u)h_j \nonumber \\
     \tr_{jk}\left[h_jT_j(u)h_kT_k(v)\right] & = & \tr_{jk}\left[T_k(v)h_kT_j(u)h_j \right] \nonumber \\
     \tr_{jk}\left[T_j(u)T_k(v)\right] & = & \tr_{jk}\left[T_k(v)T_j(u) \right] \nonumber \\
     \mathcal{T}(u)\mathcal{T}(v) & = & \mathcal{T}(v)\mathcal{T}(u).
   %  \left[\mathcal{T}(u), \mathcal{T}(v)\right] & = & 0.
\end{eqnarray}
The second and third equalities follow from the cyclicity under the partial trace and from the identity-like property of the SUSY Hamiltonian \eqref{eq:susy-h-identity-like}, respectively.

Next we construct the integrable Hamiltonian from this transfer matrix. For this we require the concept of {\it regular $R$-matrices} which is usually defined as 
\begin{eqnarray}
    R(u=0) = P~~;~~\check{R}(u=0) = \mathbb{1}.
\end{eqnarray}
It is well known that regularity of $R$-matrices act as a sufficient condition to obtain Hamiltonians that are a sum of local [nearest-neighbor interaction] terms [See \cite{slavnov2019algebraicbetheansatz} for a detailed review including a derivation]. Thus to mimic this feature for the SUSY $R$-matrices we require a notion of regularity. This is conveniently defined as follows :
\begin{eqnarray}\label{eq:R-almost-regular}
    R_{jk}(u=0) = P_{jk}h_jh_k~~;~~\check{R}_{jk}(u=0) = h_jh_k,
\end{eqnarray}
with $P$ being the usual permutation operator as defined in \eqref{eq:perm-relations}. It is clear that the operator on the right side is non-invertible due to the projectors $h$ and so we can think of this operator as a projected permutation operator. We call $R$-matrices satisfying this property as being {\it `almost' regular}. Using \eqref{eq:R-almost-regular} we can derive the expression for the Hamiltonian, defined to be, 
\begin{eqnarray}
    H :=\mathcal{T}^{-\iota}(0)\frac{d\mathcal{T}(u)}{du}\biggr|_{u=0},
\end{eqnarray}
from the transfer matrix. We will now compute each of these factors that make up the Hamiltonian. At $u=0$ the transfer matrix is evaluated to
\begin{eqnarray}
     \mathcal{T}(0)&=& tr_j[P_{j,L}h_j h_L \dots P_{j,l}h_j h_l \dots P_{j,1}h_j h_1]\nonumber\\
     &=& \left[\prod_{k=1}^L h_k\right] (P_{L,L-1} \dots P_{L,l} \dots P_{L,1})~tr_j[h_j P_{j,L} ]\nonumber\\
     &=& \left[\prod_{k=1}^L h_k\right] (P_{L,L-1} \dots P_{L,l} \dots P_{L,1}).
\end{eqnarray}
We have used the fact that $tr_j[P_{j,L}]= \mathbb{1}_L$. It is clear that $\mathcal{T}(0)$ is non-invertible due to the presence of the projectors $\left[\prod\limits_{k=1}^L h_k\right]$. Thus we can only obtain the altered inverse condition as in the case of $\check{R}$, \eqref{eq:R-inverse-altered}. This is defined as
\begin{equation}
    \mathcal{T}(0)\mathcal{T}^{-\iota}(0)=\mathcal{T}^{-\iota}(0)\mathcal{T}(0)=\prod_{k=1}^L h_k.
\end{equation}
This is satisfied when
\begin{eqnarray}
    \mathcal{T}^{-\iota}(0) = (P_{L,1} \dots P_{L,l} \dots P_{L,L-1}) \left[\prod_{k=1}^L h_k\right].
\end{eqnarray}

We will now proceed to differentiate the transfer matrix with respect to the spectral parameter $u$. To do this we first note that all the $R$-matrices we find are of the form
%\footnote{This form for the spectral parameter dependent $R$-matrix is inspired by the solutions we obtain upon Baxterizing constant invertible braid operators that will be shown in upcoming sections.}
\begin{eqnarray}
    R_{jk}(u) = P_{jk}h_jh_k + g(u)~P_{jk}\check{W}_{jk},
 \end{eqnarray}
for some function $g(u)$ that reduces to 0 for $u=0$. The operators $\check{W}_{jk}$ are constructed out of elements of the braid group algebra generated by the constant Yang-Baxter solutions $\check{Y}_{jk}$. Then we have
\begin{eqnarray}
    \frac{d\mathcal{T}(u)}{du}\biggr|_{u=0}&=& \mathfrak{c} \sum_l tr_j[P_{j,L}h_j h_L \dots P_{j,l}\check{W}_{j,l} \dots P_{j1}h_j h_1]\nonumber\\
    &=& \mathfrak{c}\left[\prod_{\substack{k=1,\\k\neq l}}^L h_k\right] \sum_l (P_{L,L-1} \dots P_{L,l}\check{W}_{N,l} \dots P_{L,1})~tr_j[h_j P_{j,L} ]\nonumber\\
    &=& \mathfrak{c}\left[\prod_{\substack{k=1,\\k\neq j}}^L h_k\right] \sum_l (P_{L,L-1} \dots P_{L,l}\check{W}_{L,l} \dots P_{L,1}).
\end{eqnarray}
Here $\mathfrak{c}=\frac{dg(u)}{du}\biggr|_{u=0}$.
Therefore, the Hamiltonian can be written as follows
\begin{equation}\label{eq:hW}
    H =\mathcal{T}^{-\iota}(0)\frac{d\mathcal{T}(u)}{du}\biggr|_{u=0}=
       \mathfrak{c} \sum_l\left[\prod_{k=1}^{l-1} h_k\right]  \check{W}_{l,l+1} \left[\prod_{k=l+2}^L h_k\right] 
\end{equation}
We have used the fact that $\check{W}_{l,l+1}h_lh_{l+1}={W}_{l,l+1}$ following the identity-like properties of the SUSY Hamiltonian \eqref{eq:susy-h-identity-like}. Note that for certain representations of the SUSY algebra the terms in $\left[\cdot\right]$ in the above expression reduce to just the identity operator in that dimension. In such cases the integrable Hamiltonian is a sum of local terms $\check{W}_{l,l+1}$. When the SUSY Hamiltonian is a non-trivial projector $h$, then the terms in $\left[ \cdot\right]$ will no longer drop out and the integrable Hamiltonian in this case is a sum of global [support on all $L$ sites] terms. However as the SUSY Hamiltonian $h$ is a sum of orthogonal projectors, one each to the `bosonic' and the `fermionic' parts of the Hilbert space, the Hamiltonian is still a sum of local terms in each of these two sectors. There are also states that entangle the two sectors and this can be seen when we consider specific models. 

With all this we have shown that the $R$-matrices constructed using SUSY algebras can lead to integrable models even if they are non-invertible as they are invertible in the altered sense \eqref{eq:R-inverse-altered}. Subsequently they lead to integrable models that is a sum of either local terms or non-local terms. 

%%%%%%%%%%%%%%%%%%%%%%%%%%%%%%%%%%%%%%%%%%%%%%%%%%%%%%%%%%%%%%%%%%%%%%
\subsection{Equivalence classes of $R$-matrices and Hamiltonian densities}
\label{subsec:equivalence-class}
%%%%%%%%%%%%%%%%%%%%%%%%%%%%%%%%%%%%%%%%%%%%%%%%%%%%%%%%%%%%%%%%%%%%%%
The final ingredient required for our work is the definition of equivalence classes of $R$-matrices and the associated Hamiltonian densities. The arguments shown here hold for both the braided and the non-braided forms of the QYBE. Given a solution of the QYBE, a potentially infinite number of other solutions can be derived from this solution with the help of the symmetries of the QYBE. Thus it is natural to expect the definition of the equivalence class to be derived from the symmetries of the quantum Yang-Baxter equation. These come in two types\footnote{There are many more symmetries of the QYBE [See for example discussions in Chapter 12 of  \cite{Essler_Frahm_Gohmann_Klümper_Korepin_2005}]. However their inclusion in the definition of an equivalence class requires a more careful study which we postpone to the future.}:
\begin{enumerate}
    \item Continuous symmetries or gauge transformations.
    \item Discrete symmetries.
\end{enumerate}
\paragraph{Continuous transformations}
For a given $R(u)$-matrix, we can generate a set of $R$-matrices
\begin{equation}\label{eq:gauge-transformation}
    R(u) \mapsto \left[Q\otimes Q\right]R(u)\left[Q\otimes Q\right]^{-1}
\end{equation}
{\it via} the single site transformation matrices $Q$. It is also easy to check that the Hamiltonian densities are related through these continuous transformations. In the case when $d=2$, we take the $Q$ matrix in the following form
\begin{equation}
    Q=\begin{pmatrix}
        g_1 & g_2\\
        g_3 & g_4
       \end{pmatrix}.
\end{equation}
We will use this notation for the $Q$ matrix while comparing Hamiltonian densities to known models in Sec.\ \ref{sec:Models-diagonalizable}.
 
\paragraph{Discrete transformations}
It is straightforward to check that if $R(u)$ represents a solution to the QYBE then, 
\begin{eqnarray}\label{eq:discrete-transformations}
    R^T(u),~~PR(u)P,~~PR^T(u)P,
\end{eqnarray}
are also solutions to the QYBE. The first two correspond to transposition and conjugation by the permutation operator respectively, while the third is a combination of the two. Discrete transformations usually lead to different Hamiltonians. However, in our case we will only consider those that are equivalent up to these transformations. For a more elaborate discussion on the equivalence classes see \cite{MSPK-Hietarinta}.

\begin{definition}
    An equivalence class of $R$-matrices comprises of elements that are related by either of the continuous or discrete symmetries of the quantum Yang-Baxter equation as listed above or a combination of these operations in any order. The same relations also define an equivalence class of local Hamiltonian densities.
\end{definition}

\begin{remark}
    When the dimension of the local Hilbert space $d$ is large, it is hard to check if two different $R$-matrices are equivalent or not. This is especially true while checking for the gauge equivalence \eqref{eq:gauge-transformation} of two $R$-matrices. However, in the SUSY construction we are guaranteed to obtain inequivalent solutions, as the supercharges which are used to construct the elements of the SUSY algebra are inequivalent. This is in fact far easier to check than checking for the equivalence of the $R$-matrices. This property can also be seen as an algebraic way of ensuring the equivalence such that it holds in all dimensions $d$. 
\end{remark}

%We are now ready to develop the different Baxterization methods of constant Yang-Baxter operators realized using elements of the $\mathcal{N}=2$ SUSY algebra.

%%%%%%%%%%%%%%%%%%%%%%%%%%%%%%%%%%%%%%%%%%%%%%%%%%%%%%%%%%%%%%%%%%%%%%
\section{Baxterization methods}
\label{sec:Baxterization}
%%%%%%%%%%%%%%%%%%%%%%%%%%%%%%%%%%%%%%%%%%%%%%%%%%%%%%%%%%%%%%%%%%%%%%
Constant invertible $4\times 4$ braid operators can be written in terms of SUSY generators using \eqref{eq:susy-representation-C2}. This results in algebraic expressions whose higher dimensional representations may or may not be invertible, depending on the representation. We will now discuss ways to Baxterize such constant Yang-Baxter solutions when they are written in terms of elements of the SUSY algebra. It will become clear that the methods presented here, in particular Methods 2, 3 and 4, generalize the traditional methods in \cite{Jones1990,li1993yang,ge-baxterization} to non-invertible constant braid operators. The methods used are all algebraic, meaning that they produce spectral parameter dependent $R$-matrices that solve the quantum Yang-Baxter equation independent of the chosen representation. Thus they yield a new solution for every representation chosen. 

We begin each case by Baxterizing the braided solutions of the Yang-Baxter equation. This yields a spectral parameter dependent $R$-matrix that satisfies the braided form of the YBE, \eqref{eq:braided-YBE}. This is then used to obtained the solution of the non-braided YBE, \eqref{eq:non-braided-YBE} by multiplying with the permutation operator. The latter is then used to construct the monodromy matrix, the transfer matrix and eventually the local Hamiltonian. 

The methods developed here depend on whether the constant Yang-Baxter solutions are diagonalizable or not. We find four methods for the former and two for the latter. We discuss all of them separately.

%%%%%%%%%%%%%%%%%%%%%%%%%%%%%%%%%%%%%%%%%%%%%%%%%%%%%%%%%%%%%%%%%%%%%%%
\subsection{Method-1: Permutation-like braid operators}
\label{subsec:method1-Baxterization}
%%%%%%%%%%%%%%%%%%%%%%%%%%%%%%%%%%%%%%%%%%%%%%%%%%%%%%%%%%%%%%%%%%%%%%%
Suppose the constant braided Yang-Baxter solution satisfies the condition
\begin{eqnarray}\label{eq:Baxterization-condition-1}
    \check{Y}_{ij}^2 =\eta~ h_ih_j~~~\eta\in\mathbb{C}.
\end{eqnarray}
Then the spectral parameter dependent solution to the braided QYBE is given by 
\begin{equation}\label{eq:ansatz-1}
    \check{R}_{ij}(u) = h_ih_j+ cu~\check{Y}_{ij}; ~~ c\in \mathbb{C}.
\end{equation}
This is almost regular \eqref{eq:R-almost-regular}, at $u=0$. The altered inverse \eqref{eq:R-inverse-altered}, of this $R$-matrix is given by
\begin{equation}
    \check{R}^{-\iota}_{ij}(u) = \frac{1}{1-\eta c^2u^2}(h_ih_j- cu~ \check{Y}_{ij}).
\end{equation}
%when we define that the $R$-matrix holds the property
%\begin{equation}
 %   R_{ij}(u)R_{ij}^{-1}(u)=h_ih_j.
%\end{equation}
Note that when $h=\mathbb{1}$, the above expression is the usual inverse.
The corresponding Hamiltonian, with a global Hamiltonian density is given by
\begin{equation}\label{eq:h-method-1}
    H =\mathcal{T}^{-\iota}(0)\frac{d\mathcal{T}(u)}{du}\biggr|_{u=0} = c \sum_j\left[\prod_{k=1}^{j-1} h_k\right]
     \check{Y}_{j,j+1}\left[\prod_{k=j+2}^L h_k\right].
\end{equation}
Comparing this with \eqref{eq:hW} we see that $\check{W}=\check{Y}$.

%%%%%%%%%%%%%%%%%%%%%%%%%%%%%%%%%%%%%%%%%%%%%%%%%%%%%%%%%%%%%%%%%%%%%%%
\subsection{Method-2:~Hecke-like braid operators}
\label{subsec:2-eigenvalues}
%%%%%%%%%%%%%%%%%%%%%%%%%%%%%%%%%%%%%%%%%%%%%%%%%%%%%%%%%%%%%%%%%%%%%%%
Next consider constant braided Yang-Baxter solutions that satisfy the Hecke-like condition
%\footnote{See Appendix \ref{app:Hecke-algebra} for the definition of the usual Iwahori-Hecke algebra. }
\begin{eqnarray}\label{eq:Baxterization-condition-hecke}
    \check{Y}_{ij}^2 =\xi~ \check{Y}_{ij}+\eta~ h_ih_j~~~\eta,\xi\in\mathbb{C}.
\end{eqnarray}
The $Y$ here satisfies an altered inverse condition as in \eqref{eq:R-inverse-altered}:
\begin{equation}
    \check{Y} \check{Y}^{-\iota}=\check{Y}^{-\iota}\check{Y}= h\otimes h.
\end{equation}
We can now construct a spectral parameter dependent solution to the braided QYBE 
\begin{equation}\label{eq:ansatz-2}
    \check{R}_{ij}(u) = h_ih_j+ \frac{(e^{c u}-1)}{\xi}~\check{Y}_{ij};~~c, \xi \in \mathbb{C}.
\end{equation}
This is also almost regular at $u=0$. The altered inverse of this $R$-matrix becomes
\begin{equation}
    \check{R}^{-\iota}_{ij}(u) = \frac{1}{1+\frac{2\eta}{\xi^2}(1-\cosh{cu})}\left[h_ih_j+ \frac{(e^{-c u}-1)}{\xi} \check{Y}_{ij}\right],
\end{equation}
The corresponding Hamiltonian is given by
\begin{equation}
    H = \mathcal{T}^{-\iota}(0)\frac{d\mathcal{T}(u)}{du}\biggr|_{u=0} = \frac{c}{\xi} \sum_j\left[\prod_{k=1}^{j-1} h_k\right]
     \check{Y}_{j,j+1}\left[\prod_{k=j+2}^L h_k\right].
\end{equation}
As in Method 1, here too we have $\check{W}=\check{Y}$ when this Hamiltonian is equated to the general form in \eqref{eq:hW}.
\begin{remark}
    Note that Method 2 is not valid when $\xi=0$. However, it is easy to see that Method 1 takes care of this situation. This observation also ensures that the two Methods produce different sets of $R$-matrices and one is not a sub-class of the other.
\end{remark}

\subsection{Method-3:~Braid operators with cubic constraints}
\label{subsec:3-eigenvalues}
%%%%%%%%%%%%%%%%%%%%%%%%%%%%%%%%%%%%%%%%%%%%%%%%%%%%%%%%%%%%%%%%%%%%%%
%Consider the constant braid solution $Y$ with three distinct eigenvalues $\lambda_1, \lambda_2, \lambda_3$. The projectors $\mathcal{P}_1, \mathcal{P}_2$ and $\mathcal{P}_3$ satisfy 
%\begin{equation}
%\mathcal{P}_1+\mathcal{P}_2+\mathcal{P}_3=(h\otimes h)~;~ \mathcal{P}_m\mathcal{P}_n=\delta_{mn}\mathcal{P}_m.
%\end{equation}
In this case, the constant solution $Y$ satisfies a cubic relation
\begin{equation}\label{eq:cubic-minimal-poly}
\check{Y}^3-(\lambda_1+\lambda_2+\lambda_3)\check{Y}^2+(\lambda_1\lambda_2+\lambda_2\lambda_3+\lambda_1\lambda_3)\check{Y}-\lambda_1\lambda_2\lambda_3~ (h\otimes h)=0,
\end{equation}
with the $\lambda_i$'s being complex parameters. The Baxterized solution of the braided YBE is then represented as 
\begin{equation} \label{eq:ansatz-3}
    \check{R}_{ij}(u)=(\lambda_1+\lambda_2+\lambda_3+\lambda_1\lambda_3\lambda_2^{-1})e^{-cu}~h_ih_j-(e^{-cu}-1)\check{Y}_{ij}+\lambda_1\lambda_3~e^{-cu}(e^{-cu}-1)\check{Y}_{ij}^{-\iota}.
\end{equation}
The altered inverse associated with the above solution, when multiplied by an overall scalar function $(1-e^{-cu})$, is given by 
\begin{equation}
    \check{R}_{ij}^{-\iota}(u)= \mathfrak{t}\left[(\lambda_1+\lambda_2+\lambda_3+\lambda_1\lambda_3\lambda_2^{-1})e^{-cu}~h_ih_j-e^{-cu}(1-e^{-cu})\check{Y}_{ij}+\lambda_1\lambda_3 (1-e^{-cu})\check{Y}_{ij}^{-\iota}\right].
\end{equation}
Here the factor $\mathfrak{t}$ is given by
$$\mathfrak{t}=\frac{1}{\left[(\lambda_2+\lambda_1\lambda_3\lambda_2^{-1})e^{-cu}+\lambda_1+\lambda_3e^{-2cu}\right]\cdot\left[(\lambda_2+\lambda_1\lambda_3\lambda_2^{-1})e^{-cu}+\lambda_3+\lambda_1e^{-2cu}\right]}.$$
The expression for the altered inverse shows that almost regular solutions can only be obtained when 
\begin{equation}\label{eq:rmatrix-constraint}
\kappa=\lambda_1+\lambda_2+\lambda_3+\lambda_1\lambda_3\lambda_2^{-1}\neq 0.
\end{equation}
Using these the local Hamiltonian corresponding to this almost regular solution is represented as 
\begin{equation}\label{eq:h-3-eigenvalues}
    H=\frac{c}{\kappa}~\sum\limits_j\left[\prod_{k=1}^{j-1} h_k\right]~\left[\check{Y}_{j,j+1}-\lambda_1\lambda_3~ \check{Y}_{j,j+1}^{-\iota}-\kappa~h_jh_{j+1}\right]~\left[\prod_{k=j+2}^{L} h_k\right]
\end{equation}
Equating this with the general form of the Hamiltonian in \eqref{eq:hW} we see that in this case
$$  \check{W}_{j,j+1}=\check{Y}_{j,j+1}-\lambda_1\lambda_3~ \check{Y}_{j,j+1}^{-\iota}-\kappa~h_jh_{j+1}. $$
It is important to highlight that the Hamiltonian is symmetric with regard to the exchange of $\lambda_1$ and $\lambda_3$. Nonetheless, the pairs $\{\lambda_1, \lambda_2\}$ and $\{\lambda_2, \lambda_3\}$, when swapped among their respective elements, yield distinct Hamiltonians.

%%%%%%%%%%%%%%%%%%%%%%%%%%%%%%%%%%%%%%%%%%%%%%%%%%%%%%%%%%%%%%%%%%%%%%
\subsection{Method-4:~Braid operators obeying a quartic constraint}
\label{subsec:4-eigenvalues}
%%%%%%%%%%%%%%%%%%%%%%%%%%%%%%%%%%%%%%%%%%%%%%%%%%%%%%%%%%%%%%%%%%%%%%
%We will now discuss about another diagonalizable solution $Y$, characterized by four distinct eigenvalues $\lambda_1, \lambda_2, \lambda_3, \lambda_4$. The projectors $\mathcal{P}_1, \mathcal{P}_2, \mathcal{P}_3$, and $\mathcal{P}_4$ satisfy 
%\begin{equation}
%\mathcal{P}_1+\mathcal{P}_2+\mathcal{P}_3+\mathcal{P}_4=(h\otimes h)~;~ \mathcal{P}_m\mathcal{P}_n=\delta_{mn}\mathcal{P}_m.
%\end{equation}
In this case, the constant solution $Y$ satisfies a quartic relation
\begin{equation}\label{eq:quartic-relation}
\check{Y}^4 -\mathfrak{d}_1 \check{Y}^3 +\mathfrak{d}_2 \check{Y}^2-\mathfrak{d}_3 \check{Y} +\mathfrak{d}_4~(h\otimes h)=0,
\end{equation}
where 
$$\mathfrak{d}_1=\sum\limits_{m=1}^4\lambda_n~,~~\mathfrak{d}_2=\sum\limits_{m<n}\lambda_m\lambda_n~,~~\mathfrak{d}_3=\sum\limits_{m<n<l}\lambda_m\lambda_n\lambda_l~,~~\mathfrak{d}_4=\prod_{m=1}^4\lambda_m,$$
are functions depending on the four complex parameters $\lambda_1$, $\lambda_2$, $\lambda_3$ and $\lambda_4$. 
The Baxterized version associated with this constant solution is 
\begin{equation}\label{eq:rmatrix-method-4}
    \check{R}_{ij}(u)=M(u)h_ih_j+L(u)\check{Y}_{ij}+K(u)\check{Y}_{ij}^2+G(u)\check{Y}_{ij}^{-\iota},
\end{equation}
where 
\begin{eqnarray}
&K(u)= e^{-cu} (e^{-cu}-1)\mathfrak{a}~,~~L(u)=(1-e^{-cu})\left[1+e^{-cu}(\mathfrak{a}\mathfrak{d}_1 +\mathfrak{b})\right]&\nonumber\\
&M(u)=e^{-cu}\left[e^{-cu}(\mathfrak{a} \mathfrak{d}_2+\lambda_1\lambda_4 \mathfrak{c})+(1+e^{-cu})\mathfrak{b}(\lambda_2+\lambda_3)+\lambda_2\lambda_3\mathfrak{c}\right]&\nonumber\\
&G(u)=e^{-cu} (e^{-cu}-1)(\lambda_1\lambda_4 e^{-cu}+\lambda_2\lambda_3 \mathfrak{b})&\nonumber
\end{eqnarray}
\begin{eqnarray}
&\mathfrak{a}=\frac{(\lambda_1\lambda_3 -\lambda_2\lambda_4 )}{\lambda_2\lambda_3(\lambda_1-\lambda_4 ) }~,~ \mathfrak{b}=\frac{\lambda_1\lambda_4(\lambda_2-\lambda_3) }{\lambda_2\lambda_3(\lambda_1-\lambda_4 )}&\nonumber\\
&\mathfrak{c}=\frac{(\lambda_1\lambda_2 -\lambda_3\lambda_4 )}{\lambda_2\lambda_3(\lambda_1-\lambda_4 ) }~,~~\mathfrak{d}_1=\sum\limits_{m=1}^4\lambda_m~,~~\mathfrak{d}_2=\sum\limits_{m<n}\lambda_m\lambda_n . &\nonumber
\end{eqnarray}
The altered inverse of this $R$-matrix is 
\begin{equation}
    \check{R}_{ij}^{-\iota}(u)=\mathfrak{t}\left[\tilde{M}(u) h_ih_j+\tilde{L}(u)\check{Y}_{ij}+\tilde{K}(u)\check{Y}_{ij}^2+\tilde{G}(u)\check{Y}_{ij}^{-\iota}\right],
\end{equation}
where 
\begin{eqnarray}
&\tilde{K}(u)= e^{-cu} (1-e^{-cu})\mathfrak{a}~,~~\tilde{L}(u)=e^{-cu}(e^{-cu}-1)\left[e^{-cu}+(\mathfrak{a}\mathfrak{d}_1 +\mathfrak{b})\right]&\nonumber\\
&\tilde{M}(u)=e^{-cu}\left[(\mathfrak{a} \mathfrak{d}_2+\lambda_1\lambda_4 \mathfrak{c})+(1+e^{-cu})\mathfrak{b}(\lambda_2+\lambda_3)+e^{-cu}\lambda_2\lambda_3\mathfrak{c}\right],&\nonumber\\
&\tilde{G}(u)=(1-e^{-cu})(\lambda_1\lambda_4+e^{-cu}\lambda_2\lambda_3 \mathfrak{b}),&\nonumber\\
&\mathfrak{t}=\frac{\lambda_2^2\lambda_3^2}{\left(\lambda _2+\lambda _1 e^{-cu}\right) \left(\lambda _1+\lambda _2 e^{-cu}\right) \left(\lambda _3+\lambda _2 e^{-cu}\right) \left(\lambda _2+\lambda _3 e^{-cu}\right) \left(\lambda _4+\lambda _3 e^{-cu}\right) \left(\lambda _3+\lambda _4 e^{-cu}\right)}.&\nonumber
\end{eqnarray}
The local Hamiltonian is then represented as 
\begin{eqnarray}\label{eq:h-method-4}
H&=&\mathfrak{m}~\sum_j\left[\prod_{k=1}^{j-1} h_k\right]~\left[(1+\mathfrak{a}\mathfrak{d}_1+\mathfrak{b})\check{Y}_{j,j+1}-(\lambda_1\lambda_4+\lambda_2\lambda_3\mathfrak{b})\check{Y}^{-\iota}_{j,j+1}-\mathfrak{a}\check{Y}^2_{j,j+1}\right.\nonumber\\&-&\left.(\lambda_2\lambda_3\mathfrak{c}+2(\mathfrak{a} \mathfrak{d}_2+\lambda_1\lambda_4 \mathfrak{c})+3\mathfrak{b}(\lambda_2+\lambda_3))h_jh_{j+1}\right]~\left[\prod_{k=j+2}^{L} h_k\right].
\end{eqnarray}
In this case the $\check{W}$ operator is the entire expression in $\left[\cdot\right]$ under the summation sign.
From this expression it is obvious that the Hamiltonian can only be obtained when the overall multiplicative scalar factor,
\begin{equation}
\mathfrak{m}=\frac{c \lambda_2\lambda_3}{\left(\lambda _1+\lambda _2\right) \left(\lambda _2+\lambda _3\right) \left(\lambda _3+\lambda _4\right)}
\end{equation}
remains finite and does not diverge. Consequently, the necessary condition for its validity is as follows:
\begin{equation}
    \left(\lambda _1+\lambda _2\right) \left(\lambda _2+\lambda _3\right) \left(\lambda _3+\lambda _4\right) \neq 0.
\end{equation}
Moreover, since not all possible choices of eigenvalues lead to valid solutions of the QYBE, it is essential to perform a direct consistency check to ensure that the QYBE \eqref{eq:braided-YBE} is satisfied.

%%%%%%%%%%%%%%%%%%%%%%%%%%%%%%%%%%%%%%%%%%%%%%%%%%%%%%%%%%%%%%%%%%%%%%%
\subsection{Method-5: Non-diagonalizable braid operators}
\label{subsec:method5-Baxterization}
%%%%%%%%%%%%%%%%%%%%%%%%%%%%%%%%%%%%%%%%%%%%%%%%%%%%%%%%%%%%%%%%%%%%%%%
We now Baxterize non-diagonalizable constant Yang-Baxter solutions. In this case the corresponding braid operator satisfies the condition:
\begin{equation}\label{eq:Baxterization condition-nilpotent}
    \check{Y}_{ij}^2=h_ih_j+ N_{ij}.
\end{equation}
where the operator $N$ is a nilpotent operator,i.e.,
\begin{equation}
N^r=0~;~~ r=2,3.
\end{equation}
The nilpotency order $r$ determines the algebraic relations satisfied by the constant braid operators. They lead to different Baxterization methods as we shall now see. In the following, we will use the shorthand notation
$$ \check{Y}_i\equiv \check{Y}_{i,i+1}~~;~~N_i\equiv N_{i, i+1},  $$
to keep the expressions simple.

%%%%%%%%%%%%%%%%%%%%%%%%%%%%%%%%%%%%%%%%%%%%%%%%%%%%%%%%%%%%%%%%%%%%%%%
\subsubsection{For $N^2=0$}
\label{subsubsec:nilpotent-1}
%%%%%%%%%%%%%%%%%%%%%%%%%%%%%%%%%%%%%%%%%%%%%%%%%%%%%%%%%%%%%%%%%%%%%%%
We require the following relations between the $\check{Y}$ and $N$ operators to be satisfied:
\begin{eqnarray}\label{eq:N-Y-relations}
    &N_i N_{i+1}=N_{i+1}N_i=0,&\nonumber\\
    &\check{Y}_{i}N_{i} =N_{i}\check{Y}_{i}=N_{i},&\nonumber\\
    &\check{Y}_{i} \check{Y}_{i+1} N_{i} = N_{i+1} \check{Y}_{i} \check{Y}_{i+1}~;~N_{i} \check{Y}_{i+1} \check{Y}_{i}= \check{Y}_{i+1} \check{Y}_{i} N_{i+1},& \nonumber \\
   & \check{Y}_{i} N_{i+1} \check{Y}_{i} =\check{Y}_{i+1} N_{i} \check{Y}_{i+1}. &
\end{eqnarray}
From \eqref{eq:Baxterization condition-nilpotent} it is clear that 
any power of the braid operator can be written as in terms of $h \otimes h$, $\check{Y}$ and $N$. This leads to the following ansatz for the $R$-matrix
\begin{equation}
    \check{R}_{ij}(u)=h_ih_{j}+ a(u)~ \check{Y}_{ij}+ b(u)~ N_{ij}.
\end{equation}
Substituting this into the braided QYBE \eqref{eq:braided-YBE} result in the functional equations:
\begin{eqnarray}\label{eq:constraints-method-5-1}
    & a(u+v)-a(u)-a(v)=0,   &  \nonumber \\
    &  b(u) = \kappa~a(u),    &  \nonumber \\
    & (2 \kappa+1) a(u) a(v)= \kappa(a(u+v)-a(u)-a(v)).      &
\end{eqnarray}
%The first equation appears as the coefficients of $\left\{\check{Y}_i, \check{Y}_{i+1}\right\}$, whereas the third one comes from the coefficient of the terms $N_{i}$ and $N_{i+1}$. The functional equation relating $a(u)$ and $b(u)$ appears in front of four terms $\check{Y}_{i}N_{i+1},~ N_{i}\check{Y}_{i+1},~ \check{Y}_{i+1}N_{i},$ $N_{i+1}\check{Y}_{i}$. 
From \eqref{eq:constraints-method-5-1}, we see that $\kappa=-\frac{1}{2}$ and $a(u)=cu$. Thus the final expression of the $R$-matrix becomes
\begin{equation}\label{eq:nilpotent-R-1}
    \check{R}_{ij}(u)= h_ih_{j}+cu \left(\check{Y}_{ij}-\frac{N_{ij}}{2}\right).
\end{equation}
The altered inverse corresponding to this solution is
\begin{equation}
     \check{R}_{ij}^{-\iota}(u)= \frac{1}{(1-c^2u^2)}\left[h_ih_{j}-cu \left(\check{Y}_{ij}-\frac{N_{ij}}{2}\right)\right],
\end{equation}
and the Hamiltonian is 
\begin{equation}\label{eq:nilpotent-h-1}
    H=c~\left[\prod_{k=1}^L h_k\right]~\sum\limits_j \left[\check{Y}_{j,j+1}-\frac{N_{j,j+1}}{2}\right].
\end{equation}
On comparing the local Hamiltonian density with that of \eqref{eq:hW}, we have $\check{W}=\check{Y}-\frac{N}{2}$.

%%%%%%%%%%%%%%%%%%%%%%%%%%%%%%%%%%%%%%%%%%%%%%%%%%%%%%%%%%%%%%%%%%%%%%%
\subsubsection{For $N^3=0$}
\label{subsubsec:nilpotent-2}
%%%%%%%%%%%%%%%%%%%%%%%%%%%%%%%%%%%%%%%%%%%%%%%%%%%%%%%%%%%%%%%%%%%%%%%
In this case the $\check{Y}$ and $N$ operators satisfy the following relations
\begin{eqnarray}
     & \check{Y}_i N_i= N_i\check{Y}_i= N_i+ \frac{N_i^2}{2}, & \nonumber \\
     &[N_i,N_{i+1}]=0~;~N_iN_{i+1}N_i = N_{i+1}N_iN_{i+1},&\nonumber\\
     %&N^2_{i} \check{Y}_{i\pm 1}N^2_{i}=0,& \nonumber\\
    & \check{Y}_i \check{Y}_{i+1} N_i=N_{i+1}\check{Y}_i\check{Y}_{i+1}~;~N_i\check{Y}_{i+1}\check{Y}_i = \check{Y}_{i+1}\check{Y}_iN_{i+1}& \nonumber\\
    & \check{Y}_i N_{i+1} \check{Y}_i=\check{Y}_{i+1}N_i \check{Y}_{i+1}& \nonumber\\
   & N^2_{i} \check{Y}_{i+1} \check{Y}_{i}= \check{Y}_{i+1} \check{Y}_{i} N^2_{i+1}~;~\check{Y}_{i} \check{Y}_{i+1} N^2_{i} = N^2_{i+1} \check{Y}_{i} \check{Y}_{i+1},& \nonumber \\
   &\check{Y}_{i} N^2_{i+1} \check{Y}_{i} =\check{Y}_{i+1} N^2_{i} \check{Y}_{i+1}~;~N_i \check{Y}_{i+1} N_i=N_{i+1}\check{Y}_iN_{i+1}. &
   \end{eqnarray}
We can write any power of this braid operator in terms of $h\otimes h$, $\check{Y}$, $N$ and $N^2$. So this time the ansatz for the $R$-matrix is a bit more complicated  
\begin{equation}
    \check{R}_{ij}(u)=h_ih_j+ a(u)~ \check{Y}_{ij}+ b(u)~ N_{ij}+c(u)~ N_{ij}^2.
\end{equation}

The coefficient corresponding to the terms $\check{Y}_i$ or $\check{Y}_{i+1}$ determines the function $a(u)$ to be linear. The coefficients of $N_i\check{Y}_{i+1}$, $N_{i+1}\check{Y}_i$, $\check{Y}_iN_{i+1}$, $\check{Y}_{i+1}N_i$ and $N^2_i\check{Y}_{i+1}$, $N^2_{i+1}\check{Y}_i$, $\check{Y}_iN^2_{i+1}$, $\check{Y}_{i+1}N^2_i$ result in the value of $b(u)$ and $c(u)$ 
\begin{equation}
b(u)= \nu_1~a(u)~;~ c(u)= \nu_2~a(u),
\end{equation}
respectively.
The constants $\nu_1=-\frac{1}{2}, \nu_2=\frac{1}{8}$ are determined by substituting the functions $b(u)$ and $c(u)$ into the coefficients of $N_i$ and $N_i^2$ respectively. 
These are given by 
\begin{eqnarray}
    & b(u)+b(v)+a(u)b(v)+b(u)a(v)+a(u)a(v)-b(u+v)=0   ,& \\
    &  c(u)+c(v)+b(u)b(v)+a(u)c(v)+a(v)c(u)+\frac{(a(u)b(v)+b(u)a(v))}{2}=c(u+v),& 
\end{eqnarray}
respectively. Solving these functional equations lead to the final expression for the $R$-matrix as  
\begin{equation}\label{eq:nilpotent-R-2}
    \check{R}_{ij}(u)=h_i h_j+ cu\left( \check{Y}_{ij}-\frac{1}{2} N_{ij}+\frac{1}{8} N_{ij}^2\right).
\end{equation}
The altered inverse of this $R$-matrix is 
\begin{equation}
     \check{R}_{ij}^{-\iota}(u)= \frac{1}{(1-c^2u^2)}\left[h_ih_j-cu \left(\check{Y}_{ij}-\frac{N_{ij}}{2}+\frac{N^2_{ij}}{8}\right)\right].
\end{equation}
The expression of the local Hamiltonian is
\begin{equation}\label{eq:nilpotent-h-2}
    H=c~\left[\prod_{k=1}^L h_k\right]~\sum\limits_j \left[\check{Y}_{j,j+1}-\frac{N_{j,j+1}}{2}+\frac{N^2_{j,j+1}}{8}\right].
\end{equation}
The Hamiltonian density compared with \eqref{eq:hW} shows that 
$$ \check{W} = \check{Y} -\frac{N}{2} + \frac{N^2}{8}.$$

%%%%%%%%%%%%%%%%%%%%%%%%%%%%%%%%%%%%%%%%%%%%%%%%%%%%%%%%%%%%%%%%%%%%%%
\section{Hamiltonians}
\label{sec:hamiltonians}
%%%%%%%%%%%%%%%%%%%%%%%%%%%%%%%%%%%%%%%%%%%%%%%%%%%%%%%%%%%%%%%%%%%%%%
The constant $4\times 4$ solutions of QYBE were classified in a series of papers in the early 90's \cite{HIETARINTA-PLA,hietarinta1993-JMP-Long}.  The classification includes 10 invertible equivalence classes. We denote these classes by $Hx,y$. The integer $x$ denotes the number of independent parameters describing the elements of that class. The integer $y$ is a serial number indexing the number of the solution with $x$ independent parameters. So $Hx,y$ denotes the $y$th solution with $x$ independent parameters. 

We find that eight of the ten invertible classes, $H1,1$, $H1,2$, $H1,3$, $H1,4$, $H2,1$, $H2,2$, $H0,2$, and $H3,1$,are diagonalizable and two, $H0,1$ and $H2,3$, are non-diagonalizable. They can be Baxterized using one of the five techniques discussed in Sec.\ \ref{sec:Baxterization}. For each case we carry out the following steps :
\begin{enumerate}
    \item We first write down a representative of the equivalence class in the $\mathbb{C}^2$ representation i.e. a $4\times 4$ matrix.
    \item Then using the isomorphism between the $\mathcal{N}=2$ SUSY algebra and ${\mathcal Mat}(2, \mathbb{C})$ \eqref{eq:susy-representation-C2}, we can write down the algebraic expressions for these $4\times 4$ braid operators in terms of SUSY algebra elements.
    \item We then specify the Baxterization method [Sec.\ \ref{sec:Baxterization}] that applies for this class, leading to the representation independent $R$-matrix and the integrable Hamiltonian. 
    \item Then choosing the $\mathbb{C}^2$ representation, the $4\times 4$ version of the $R$-matrix and the Hamiltonian density are written down. They are then compared with known Hamiltonians in the literature.
\end{enumerate}

%We will now demonstrate this procedure for each class separately.

%%%%%%%%%%%%%%%%%%%%%%%%%%%%%%%%%%%%%%%%%%%%%%%%%%%%%%%%%%%%%%%%%%%%%%
\subsection{Diagonalizable classes}
\label{sec:Models-diagonalizable}
%%%%%%%%%%%%%%%%%%%%%%%%%%%%%%%%%%%%%%%%%%%%%%%%%%%%%%%%%%%%%%%%%%%%%%
We begin with the diagonalizable cases. Each of the eight diagonalizable Hietarinta classes are considered separately. We will use Baxterization Methods in Secs. \ref{subsec:method1-Baxterization}-\ref{subsec:4-eigenvalues} to obtain the Hamiltonians.

%%%%%%%%%%%%%%%%%%%%%%%%%%%%%%
\subsubsection{H1,1 class:} 
\label{subsec:H11}
%%%%%%%%%%%%%%%%%%%%%%%%%%%%%%%
\begin{itemize}
    \item The $4 \times 4$ representation of the braid operator for this class is
    \begin{equation}
      \check{Y}^{(1,1)} =   \begin{pmatrix}
 \beta ^2+2 \beta  \gamma -\gamma ^2 & 0 & 0 & \beta ^2-\gamma ^2 \\
 0 & \beta ^2-\gamma ^2 & \beta ^2+\gamma ^2 & 0 \\
 0 & \beta ^2+\gamma ^2 & \beta ^2-\gamma ^2 & 0 \\
 \beta ^2-\gamma ^2 & 0 & 0 & \beta ^2-2 \beta  \gamma -\gamma ^2 \\
        \end{pmatrix}.
    \end{equation}
    \item The algebraic expression for this class using SUSY generators is
\begin{eqnarray}\label{eq:yc11}
    \check{Y}^{(1,1)}_{ij} &=& (\beta^2+2 \beta \gamma-\gamma^2) b_i b_j+(\beta^2-2 \beta \gamma-\gamma^2) f_i f_j \nonumber \\ & + &(\beta^2-\gamma^2)(b_i f_j+f_i b_j+q_i q_j+q^\dag_i q^\dag_j)
    + (\beta^2+\gamma^2)(q_i q^\dag_j+ q^\dag_i q_j).
\end{eqnarray}
\item This can be Baxterized using Method 2 in Sec.\ \ref{sec:Baxterization} as it satisfies the Hecke-like condition $$\check{Y}_{ij}^2 =\xi~ \check{Y}_{ij}+\eta~ h_ih_j,~~\xi= 2 (\beta^2-\gamma^2)~,~ \eta=4 \beta^2\gamma^2.$$
\item The algebraic expression of the $R$-matrix is 
\begin{equation}\label{eq:R-matrix-H11}
    \check{R}^{(1,1)}_{ij}(u) = h_ih_j+ \frac{(e^{c u}-1)}{ 2 (\beta^2-\gamma^2)}~\check{Y}^{(1,1)}_{ij};~~c, \beta, \gamma \in \mathbb{C}.
\end{equation}
\item The algebraic expression for the integrable model is
\begin{equation}\label{eq:hdensity-H11}
H = \frac{c}{2 (\beta^2-\gamma^2)}~\left[\prod_{k=1}^{L} h_k\right]~ \sum \limits_j \check{Y}^{(1,1)}_{j,j+1}.
\end{equation}
\item In the $\mathbb{C}^2$ representation, the $R$-matrix will be
\begin{equation}
\check{R}^{(1,1)}(u) =
\begin{pmatrix}
 \frac{\left(\beta ^2+2 \beta  \gamma -\gamma ^2\right) \left(e^{c u}-1\right)}{2 \left(\beta ^2-\gamma ^2\right)}+1 & 0 & 0 & \frac{1}{2} \left(e^{c u}-1\right) \\
 0 & \frac{1}{2} \left(e^{c u}+1\right) & \frac{\left(\beta ^2+\gamma ^2\right) \left(e^{c u}-1\right)}{2 \left(\beta ^2-\gamma ^2\right)} & 0 \\
 0 & \frac{\left(\beta ^2+\gamma ^2\right) \left(e^{c u}-1\right)}{2 \left(\beta ^2-\gamma ^2\right)} & \frac{1}{2} \left(e^{c u}+1\right) & 0 \\
 \frac{1}{2} \left(e^{c u}-1\right) & 0 & 0 & \frac{\left(\beta ^2-2 \beta  \gamma -\gamma ^2\right) \left(e^{c u}-1\right)}{2 \left(\beta ^2-\gamma ^2\right)}+1 \\
\end{pmatrix}.
\end{equation}
\item In the $\mathbb{C}^2$ representation, $h = \mathbb{1}$ and so the local Hamiltonian density in $\mathbb{C}^2 \otimes \mathbb{C}^2$ can be obtained by substituting \eqref{eq:q2sigma} in \eqref{eq:yc11} and \eqref{eq:hdensity-H11}, 
\begin{eqnarray}
 \mathfrak{h}_{(1,1)} %(\beta^2-\gamma^2)\mathbb{1}+
&=&\beta^2~ \sigma^x_j\sigma^x_{j+1}+\gamma^2~\sigma^y_j\sigma^y_{j+1}+\beta\gamma~(\sigma^z_j+\sigma^z_{j+1}).
\end{eqnarray}
This falls into the family of eight vertex models \cite{vieira2018solving, Sogo}.
%\begin{equation}
%\mathfrak{h}_{(1,1)}=\kappa\left[Q\otimes Q\right]\begin{pmatrix}
 %a_1 & 0 & 0 & d_1 \\
 %0 & s_1 & c_1 & 0 \\
 %0 & c_1 & s_1 & 0 \\
 %d_2 & 0 & 0 & 2 s_1-a_1 \\
 %\end{pmatrix}\left[Q\otimes Q\right]^{-1},
%\end{equation}
%with the gauge transformations
%\begin{eqnarray*}
 %   &\left\{g_2\to 0,g_3\to 0,a_1\to \frac{\beta ^2+2 \beta  \gamma -\gamma ^2}{\kappa },s_1\to \frac{(\beta^2 -\gamma^2 )}{\kappa },c_1\to \frac{\beta ^2+\gamma ^2}{\kappa }\right\},&\\&\left\{d_1\to \frac{g_4^2 (\beta^2 -\gamma^2 )}{g_1^2 \kappa },d_2\to \frac{g_1^2 (\beta^2 -\gamma^2 )}{g_4^2 \kappa }\right\}.&
%\end{eqnarray*}
\end{itemize}

%%%%%%%%%%%%%%%%%%%%%%%%%%%%%%%%%%
\subsubsection{H1,2 class:}
\label{subsec:H12}
%%%%%%%%%%%%%%%%%%%%%%%%%%%%%%%%%%
\begin{itemize}
    \item The $4 \times 4$ representative of the braid operator is
   \begin{equation}
       \check{Y}^{(1,2)} =  \begin{pmatrix}
 \beta  & 0 & 0 & \alpha  \\
 0 & \beta -\gamma  & \beta  & 0 \\
 0 & \gamma  & 0 & 0 \\
 0 & 0 & 0 & -\gamma  \\
        \end{pmatrix}.
    \end{equation}
    \item The SUSY realization can be represented as
    \begin{equation}
         \check{Y}^{(1,2)}_{ij}= \beta~( b_ib_j+q_iq^\dag_j)+\gamma~(q^\dag_iq_j-f_if_j)+ (\beta-\gamma)b_if_j+\alpha~ q_iq_j
    \end{equation}
    \item It can be Baxterized by Method 2 of Sec.\ \ref{sec:Baxterization} as it satisfies $\check{Y}_{ij}^2 =\xi~ \check{Y}_{ij}+\eta~ h_ih_j$, with $\xi=  (\beta-\gamma)~,~ \eta=\beta\gamma$.
    \item The algebraic expression of the $R$-matrix is 
\begin{equation}\label{eq:R-matrix-H12}
    \check{R}^{(1,2)}_{ij}(u) = h_ih_j+ \frac{(e^{c u}-1)}{(\beta-\gamma)}~\check{Y}^{(1,2)}_{ij};~~c, \beta, \gamma \in \mathbb{C}.
\end{equation}
    \item The algebraic expression for the Hamiltonian is
\begin{equation}
H = \frac{c}{ (\beta-\gamma)}~\left[\prod_{k=1}^{L} h_k\right]~ \sum \limits_j \check{Y}^{(1,2)}_{j,j+1}.
\end{equation}
\item In the $\mathbb{C}^2$ representation, the $R$-matrix becomes
\begin{equation}
\check{R}^{(1,2)}(u)=
\begin{pmatrix}
 \frac{\beta  \left(e^{c u}-1\right)}{\beta -\gamma }+1 & 0 & 0 & \frac{\alpha  \left(e^{c u}-1\right)}{\beta -\gamma } \\
 0 & e^{c u} & \frac{\beta  \left(e^{c u}-1\right)}{\beta -\gamma } & 0 \\
 0 & \frac{\gamma  \left(e^{c u}-1\right)}{\beta -\gamma } & 1 & 0 \\
 0 & 0 & 0 & \frac{\beta -\gamma  e^{c u}}{\beta -\gamma } \\
\end{pmatrix}.
\end{equation}
\item The local Hamiltonian density acting on $\mathbb{C}^2 \otimes \mathbb{C}^2$ is
\begin{eqnarray}
\mathfrak{h}_{(1,2)}
% &=& \frac{(\beta-\gamma)}{2}\mathbb{1}+\frac{(\alpha+\beta+\gamma)}{4}\sigma^x_j\sigma^x_{j+1}+\frac{(\beta+\gamma-\alpha)}{4}\sigma^y_j\sigma^y_{j+1}\nonumber \\ & + & \frac{\mathrm{i}(\alpha-\beta+\gamma)}{4}\sigma^x_j\sigma^y_{j+1}+\frac{\mathrm{i}(\alpha+\beta-\gamma)}{4}\sigma^y_j\sigma^x_{j+1}+\frac{1}{2}(\beta~\sigma^z_j+\gamma~\sigma^z_{i+1}).
 =  \alpha~\sigma_j^+\sigma_{j+1}^+ + \beta~\sigma_j^+\sigma_{j+1}^- + \gamma~\sigma^-_j\sigma^+_{j+1} +   \frac{1}{2}\left[\beta~\sigma^z_j+\gamma~\sigma^z_{i+1}\right].
\end{eqnarray}
This is gauge equivalent to the seven-vertex type model.
%\begin{equation}
 % \mathfrak{h}_{(1,2)}=\kappa\left[Q\otimes Q\right]  \begin{pmatrix}
 %a_1 & 0 & 0 & d_1 \\
 %0 & a_1-c_2 & c_1 & 0 \\
 %0 & c_2 & a_1-c_1 & 0 \\
 %0 & 0 & 0 & a_1-c_1-c_2 \\
  %  \end{pmatrix}\left[Q\otimes Q\right]^{-1},
%\end{equation}
%with the following transformations
%$$\left\{g_2\to 0,g_3\to 0,a_1\to \frac{\beta }{\kappa },c_1\to \frac{\beta }{\kappa },d_1\to \frac{\alpha  g_4^2}{g_1^2 \kappa },c_2\to \frac{\gamma }{\kappa }\right\}.$$
\end{itemize}

%%%%%%%%%%%%%%%%%%%%%%%%%%%%%%
\subsubsection{H1,3 class:}
\label{subsec:H13}
%%%%%%%%%%%%%%%%%%%%%%%%%%%%%%
\begin{itemize}
    \item The $4 \times 4$ representative is
    \begin{equation}
       \check{Y}^{(1,3)} = \begin{pmatrix}
 \alpha ^2 & -\alpha  \beta  & \alpha  \beta  & \beta  \gamma  \\
 0 & 0 & \alpha ^2 & \alpha  \gamma  \\
 0 & \alpha ^2 & 0 & -\alpha  \gamma  \\
 0 & 0 & 0 & \alpha ^2 \\
        \end{pmatrix}.
    \end{equation}
    \item In terms of SUSY generators this becomes
    \begin{equation}
         \check{Y}^{(1,3)}_{ij}= \alpha^2~( b_ib_j+f_if_j+q^\dag_iq_j+q_iq^\dag_j)+\alpha\beta~(q_ib_j-b_iq_j)+ \alpha\gamma~ (q_if_j- f_iq_j)+\beta \gamma~q_iq_j.
    \end{equation}
    \item It is Baxterized by Method 1 of Sec.\ \ref{sec:Baxterization} as it  satisfies the condition $\check{Y}_{ij}^2 =\eta~ h_ih_j$, with $\eta=\alpha^4$.
    \item The algebraic forms of the $R$-matrix and the Hamiltonian are expressed as
    \begin{eqnarray}\label{eq:R-matrix-H13}
    \check{R}^{(1,3)}_{ij}(u) & = & h_ih_j+ cu~\check{Y}^{(1,3)}_{ij}; ~~ c\in \mathbb{C},  \\
    H & = & c~\left[\prod_{k=1}^{L} h_k\right]~ \sum \limits_j \check{Y}^{(1,3)}_{j,j+1}.
    \end{eqnarray}
    \item In the $\mathbb{C}^2$ representation, the $R$-matrix will be
\begin{equation}
\check{R}^{(1,3)}(u)=
\begin{pmatrix}
 \alpha ^2 c u+1 & -c u\alpha  \beta  & c u\alpha  \beta  & c u\beta \gamma \\
 0 & 1 & \alpha ^2 c u & cu\alpha \gamma  \\
 0 & \alpha ^2 c u & 1 & -cu\alpha \gamma  \\
 0 & 0 & 0 & \alpha ^2 c u+1 \\
\end{pmatrix}.
\end{equation}
    \item In $\mathbb{C}^2$, the local Hamiltonian corresponding to this $R$-matrix is given by 
    \begin{eqnarray} \label{eq:hdensity-H13}
    \mathfrak{h}_{(1,3)}=\alpha^2~P_{j,j+1}+\beta\gamma~\sigma^+_j\sigma^+_{j+1}+\frac{\alpha(\beta-\gamma)}{2}(\sigma^+_j\sigma^z_{j+1}-\sigma^z_j\sigma^+_{j+1}).
     %\mathfrak{h}_{(1,3)}
      %  &=& \frac{\alpha^2}{2}(\mathbb{1}+\sigma^z_j\sigma^z_{j+1})+\frac{(2\alpha^2+\beta\gamma)}{4}\sigma^x_j\sigma^x_{j+1}+\frac{(2\alpha^2-\beta\gamma)}{4}\sigma^y_j\sigma^y_{j+1}\nonumber\\&+&\frac{\mathrm{i}\beta\gamma}{4}(\sigma^x_j\sigma^y_{j+1}+\sigma^y_j\sigma^x_{j+1})+\frac{\alpha(\beta-\gamma)}{2}(\sigma^+_j\sigma^z_{j+1}-\sigma^z_j\sigma^+_{j+1})\nonumber\\&+&\frac{\alpha(\beta+\gamma)}{2}(\sigma^+_j-\sigma^+_{j+1}).
    \end{eqnarray}
Here $\sigma^{\pm}=\frac{\sigma^x\pm \mathrm{i}\sigma^y}{2}$ and $P$ is the standard permutation operator on $\mathbb{C}^2\otimes\mathbb{C}^2$. An important observation is that this local Hamiltonian density does not fit into any of the current classifications of nearest-neighbor spin $\frac{1}{2}$ Hamiltonians. 
\begin{remark}
The resulting system is a non-Hermitian deformation of the permutation operator. It preserves the permutation relations \eqref{eq:perm-relations}, but does not exhibit the exchange property. It is then worth noting that a sub-class of the braid operator in the $H1,3$ class can be related to the usual permutation operator \textit{via} a twist transformation (See Lemma 11, Chapter 12 of \cite{Essler_Frahm_Gohmann_Klümper_Korepin_2005}). This is achieved with a single-site twist element $A$, acting on the braid operator, $\check{Y}$ as
\begin{equation}
    P_{ij}= A_i \check{Y}_{ij} A_i^{-1}~;~A=\begin{pmatrix}
        1 & \beta \\
        0 & 1
    \end{pmatrix},
\end{equation}
with the special choice $\gamma=-\beta$ and $\alpha$, scaled to 1 for simplicity. As opposed to the local gauge transformations, \eqref{eq:gauge-transformation}, the twist transformation acts on only one of the indices of the braid operator. Nevertheless, it is easy to see that it can still satisfy the braided QYBE when
\begin{equation}
    [A\otimes A, \check{Y}]=0.
\end{equation}
This is indeed true for the above example, as can be easily verified.
%It means that under this condition, the twisted braid operator continues to satisfy the same braided QYBE as the original operator $\check{Y}$. 
As the Hamiltonian is constructed as the sum of the local $\check{Y}$'s, we have a consistent effect on the Hamiltonian {\it via} a similar twist transformation. Thus the Hamiltonian density is connected to the permutation operator through a similar unipotent twist element. 
It is also clear that the non-hermitian deformations present in the Hamiltonian density remains unaltered under the twist, though the coefficients involved would be modified.
\end{remark}
\end{itemize}

%%%%%%%%%%%%%%%%%%%%%%%%%%%%%%%%
\subsubsection{H2,1 class:}
\label{subsec:H21}
%%%%%%%%%%%%%%%%%%%%%%%%%%%%%%%%
\begin{itemize}
    \item We begin with the $4 \times 4$ form of the braid operator
    \begin{equation}
       \check{Y}^{(2,1)}= \begin{pmatrix}
 \alpha ^2 & 0 & 0 & 0 \\
 0 & \alpha ^2-\beta  \gamma  & \alpha  \beta  & 0 \\
 0 & \alpha  \gamma  & 0 & 0 \\
 0 & 0 & 0 & \alpha ^2 \\
        \end{pmatrix}.
    \end{equation}
    \item In terms of the SUSY generators this is
    \begin{equation}
         \check{Y}^{(2,1)}_{ij}= \alpha^2~( b_ib_j+f_if_j)+(\alpha^2-\beta\gamma)~b_if_j+ \alpha\beta~ q_iq^\dag_j+\alpha \gamma~q^\dag_iq_j.
    \end{equation}
    \item This is Baxterized by Method 2 of Sec.\ \ref{sec:Baxterization} : $\check{Y}_{ij}^2 =\xi~ \check{Y}_{ij}+\eta~ h_ih_j$, where $\xi=  (\alpha^2-\beta\gamma)~,~ \eta=\alpha^2\beta\gamma$.
    \item The algebraic expression of the $R$-matrix is 
\begin{equation}\label{eq:R-matrix-H21}
    \check{R}^{(2,1)}_{ij}(u) = h_ih_j+ \frac{(e^{c u}-1)}{(\alpha^2-\beta\gamma)}~\check{Y}^{(2,1)}_{ij};~~c, \alpha, \beta, \gamma \in \mathbb{C}.
\end{equation}
    \item The algebraic expression for the Hamiltonian is 
    \begin{equation}
        H = \frac{c}{ (\alpha^2-\beta\gamma)}~\left[\prod_{k=1}^{L} h_k\right]~\sum \limits_j \check{Y}^{(2,1)}_{j,j+1}.
    \end{equation}
    \item In the $\mathbb{C}^2$ representation, the $R$-matrix is
\begin{equation}
\check{R}^{(2,1)}(u)=
\begin{pmatrix}
 \frac{\alpha ^2 \left(e^{c u}-1\right)}{\alpha ^2-\beta  \gamma }+1 & 0 & 0 & 0 \\
 0 & e^{c u} & \frac{\alpha  \beta  \left(e^{c u}-1\right)}{\alpha ^2-\beta  \gamma } & 0 \\
 0 & \frac{\alpha  \gamma  \left(e^{c u}-1\right)}{\alpha ^2-\beta  \gamma } & 1 & 0 \\
 0 & 0 & 0 & \frac{\alpha ^2 \left(e^{c u}-1\right)}{\alpha ^2-\beta  \gamma }+1 \\
\end{pmatrix}.
\end{equation}
    \item In the $\mathbb{C}^2$ representation, the local term of the Hamiltonian looks like 
    \begin{eqnarray}
     \mathfrak{h}_{(2,1)}
%        &=& \frac{(3\alpha^2-\beta\gamma)}{4}\mathbb{1}+\frac{\alpha(\beta+\gamma)}{4}(\sigma^x_j\sigma^x_{j+1}+\sigma^y_j\sigma^y_{j+1})+\frac{(\alpha^2+\beta\gamma)}{4}\sigma^z_j\sigma^z_{j+1}\nonumber\\&+&\frac{\mathrm{i}\alpha(\beta-\gamma)}{4}(\sigma^y_j\sigma^x_{j+1}-\sigma^x_j\sigma^y_{j+1})+\frac{(\alpha^2-\beta\gamma)}{4}(\sigma^z_j-\sigma^z_{j+1}),
=\alpha\beta~\sigma_j^+\sigma_{j+1}^- + \alpha\gamma~\sigma_j^-\sigma_{j+1}^+ + \frac{(\alpha^2+\beta\gamma)}{4}\sigma^z_j\sigma^z_{j+1}.
    \end{eqnarray}
%The last term in the expression represents an external magnetic field, which vanishes in the periodic spin chain. 
This belongs to the $XXZ$ family. 
%\begin{equation}\label{eq:xxz-model-1}
 % \mathfrak{h}_{(2,1)}=\kappa\left[Q\otimes Q\right]   \begin{pmatrix}
 %a_1 & 0 & 0 & 0 \\
 %0 & s_1 & c_1 & 0 \\
 %0 & c_2 & s_2 & 0 \\
 %0 & 0 & 0 & a_1 \\
  %  \end{pmatrix}\left[Q\otimes Q\right]^{-1},
%\end{equation}
%{\it via} the following gauge transformations 
%$$\left\{g_1\to 0,g_4\to 0,a_1\to \frac{\alpha ^2}{\kappa },s_1\to 0,c_1\to \frac{\alpha  \gamma }{\kappa },s_2\to \frac{\alpha ^2-\beta  \gamma }{\kappa },c_2\to \frac{\alpha  \beta }{\kappa }\right\}.$$
\end{itemize}

%%%%%%%%%%%%%%%%%%%%%%%%%%%%%%%
\subsubsection{H2,2 class:}
\label{subsec:H22}
%%%%%%%%%%%%%%%%%%%%%%%%%%%%%%%
\begin{itemize}
    \item The $4\times 4$ braid operator is represented as 
    \begin{equation}
       \check{Y}^{(2,2)}= \begin{pmatrix}
 \alpha ^2 & 0 & 0 & 0 \\
 0 & \alpha ^2-\beta  \gamma  & \alpha  \beta  & 0 \\
 0 & \alpha  \gamma  & 0 & 0 \\
 0 & 0 & 0 & -\beta  \gamma  \\
        \end{pmatrix}.
    \end{equation}
    \item The SUSY version is
    \begin{equation}
         \check{Y}^{(2,2)}_{ij}= \alpha^2~ b_ib_j-\beta\gamma f_if_j+(\alpha^2-\beta\gamma)~b_if_j+ \alpha\beta~ q_iq^\dag_j+\alpha \gamma~q^\dag_iq_j.
    \end{equation}
    \item It is Baxterized by Method 2 of Sec.\ \ref{sec:Baxterization}, as it satisfies $\check{Y}_{ij}^2 =\xi~ \check{Y}_{ij}+\eta~ h_ih_j$, with $\xi=  (\alpha^2-\beta\gamma)~,~ \eta=\alpha^2\beta\gamma$.
    \item The algebraic expression of the $R$-matrix is 
\begin{equation}\label{eq:R-matrix-H22}
    \check{R}^{(2,2)}_{ij}(u) = h_ih_j+ \frac{(e^{c u}-1)}{(\alpha^2-\beta\gamma)}~\check{Y}^{(2,2)}_{ij};~~c, \alpha, \beta, \gamma \in \mathbb{C}.
\end{equation}
    \item The integrable Hamiltonian is 
    \begin{equation}
        H = \frac{c}{ (\alpha^2-\beta\gamma)}~\left[\prod_{k=1}^{L} h_k\right]~\sum \limits_j \check{Y}^{(2,2)}_{j,j+1}.
    \end{equation}
    \item The $4\times 4$ $R$-matrix will be
\begin{equation}
\check{R}^{(2,2)}(u)=
\begin{pmatrix}
 \frac{\alpha ^2 \left(e^{c u}-1\right)}{\alpha ^2-\beta  \gamma }+1 & 0 & 0 & 0 \\
 0 & e^{c u} & \frac{\alpha  \beta  \left(e^{c u}-1\right)}{\alpha ^2-\beta  \gamma } & 0 \\
 0 & \frac{\alpha  \gamma  \left(e^{c u}-1\right)}{\alpha ^2-\beta  \gamma } & 1 & 0 \\
 0 & 0 & 0 & \frac{\beta  \gamma  \left(e^{c u}-1\right)}{\beta  \gamma -\alpha ^2}+1 \\
\end{pmatrix}.
\end{equation}
    \item And on $\mathbb{C}^2\otimes\mathbb{C}^2$, the Hamiltonian density becomes
    \begin{eqnarray}
     \mathfrak{h}_{(2,2)}
%    &=& \frac{(\alpha^2-\beta\gamma)}{2}\mathbb{1}+\frac{\alpha(\beta+\gamma)}{4}(\sigma^x_j\sigma^x_{j+1}+\sigma^y_j\sigma^y_{j+1}) \nonumber \\ &+& \frac{\mathrm{i}\alpha(\beta-\gamma)}{4}(\sigma^y_j\sigma^x_{j+1}-\sigma^x_j\sigma^y_{j+1}) + \frac{\alpha^2}{4}\sigma^z_j+\frac{\beta\gamma}{2}\sigma^z_{j+1}.
= \alpha\beta~\sigma^+_j\sigma^-_{j+1} + \alpha\gamma~\sigma^-_j\sigma^+_{j+1} + \frac{\alpha^2}{4}\sigma^z_j+\frac{\beta\gamma}{2}\sigma^z_{j+1}.
    \end{eqnarray}
It is important to note that this local Hamiltonian density does not belong to any of the existing classes of the Hamiltonians.
\end{itemize}
%%%%%%%%%%%%%%%%%%%%%%%%%%%%%%%%%%%%%
\subsubsection{H0,2 class:}
\label{subsec:H02}
%%%%%%%%%%%%%%%%%%%%%%%%%%%%%%%%%%%%%
\begin{itemize}
    \item The $4 \times 4$ representative of the braid operator for this class is expressed as
    \begin{equation}
       \check{Y}^{(0,2)}= P(P R_{H0,2} P) =\begin{pmatrix}
 1 & 0 & 0 & 1 \\
 0 & 1 & 1 & 0 \\
 0 & -1 & 1 & 0 \\
 -1 & 0 & 0 & 1 \\
        \end{pmatrix},
    \end{equation}
    where $R_{H0,2}$ is the operator found in \cite{HIETARINTA-PLA}.
    \item In terms of the SUSY generators it becomes
    \begin{equation}
         \check{Y}^{(0,2)}_{ij}= ( b_ib_j+f_if_j+b_if_j+f_ib_j)+(q_iq_j+ q_iq^\dag_j)-(q^\dag_iq_j+q^\dag_iq^\dag_j).
    \end{equation}
    \item This too is Baxterized by method 2, as $\check{Y}_{ij}^2 =\xi~ \check{Y}_{ij}+\eta~ h_ih_j$, with $\xi= 2~,~ \eta=-2$.
    \item The algebraic expression of the $R$-matrix is 
\begin{equation}\label{eq:R-matrix-H02}
    \check{R}^{(0,2)}_{ij}(u) = h_ih_j+ \frac{(e^{c u}-1)}{2}~\check{Y}^{(0,2)}_{ij};~~c, \alpha, \beta, \gamma \in \mathbb{C}.
\end{equation}
    \item The algebraic form of the Hamiltonian is
    \begin{equation}
        H = \frac{c}{2}~\left[\prod_{k=1}^{L} h_k\right]~\sum \limits_j \check{Y} ^{(0,2)}_{j,j+1}.
    \end{equation}
     \item The $4\times 4$ form of the $R$-matrix is
\begin{equation}
\check{R}^{(0,2)}(u)=\frac{1}{2}
\begin{pmatrix}
 e^{c u}+1 & 0 & 0 & e^{c u}-1 \\
 0 & e^{c u}+1 & e^{c u}-1 & 0 \\
 0 & 1-e^{c u} & e^{c u}+1 & 0 \\
 1-e^{c u} & 0 & 0 & e^{c u}+1 \\
\end{pmatrix}.
\end{equation}
   \item The corresponding local term of the Hamiltonian in $\mathbb{C}^2$ looks like 
    \begin{eqnarray}
     \mathfrak{h}_{(0,2)}&=& \mathbb{1}+\mathrm{i}~\sigma^y_j\sigma^x_{j+1},
    \end{eqnarray}
which belongs to one of the eight-vertex [$XYZ$] families. 
%\begin{equation}
 %  \mathfrak{h}_{(0,2)}=\kappa\left[Q\otimes Q\right] \begin{pmatrix}
 %a_1 & 0 & 0 & d_1 \\
 %0 & a_1 & s_1 & 0 \\
 %0 & -s_1 & a_1 & 0 \\
 %d_2 & 0 & 0 & a_1 \\
 %   \end{pmatrix}\left[Q\otimes Q\right]^{-1},
%\end{equation}
%through the following gauge transformations 
%$$\left\{g_2\to 0,g_3\to 0,a_1\to \frac{1}{\kappa },s_1\to \frac{1}{\kappa },d_1\to \frac{g_4^2}{g_1^2 \kappa },d_2\to -\frac{g_1^2}{g_4^2 \kappa }\right\}.$$
\end{itemize}

%%%%%%%%%%%%%%%%%%%%%%%%%%%%%%%%%%%%%
\subsubsection{H1,4 class:}
\label{subsec:H14}
%%%%%%%%%%%%%%%%%%%%%%%%%%%%%%%%%%%%%
\begin{itemize}
    \item The $4\times 4$ representative is given by 
    \begin{eqnarray}
       \check{Y}^{(1,4)} =   \begin{pmatrix}
 0 & 0 & 0 & \alpha  \\
 0 & 1 & 0 & 0 \\
 0 & 0 & 1 & 0 \\
 \beta  & 0 & 0 & 0 \\
    \end{pmatrix}.
    \end{eqnarray}
    \item The SUSY versions for this operator and its altered inverse are
    \begin{eqnarray}
         \check{Y}_{ij}^{(1,4)} & = & \alpha~ q_iq_j+\beta~ q^\dag_i q^\dag_j+b_if_j+f_ib_j, \\ \nonumber 
         \check{Y}_{ij}^{(1,4)-\iota} & = & \frac{1}{\beta}~ q_iq_j+\frac{1}{\alpha}~ q^\dag_i q^\dag_j+b_if_j+f_ib_j. \end{eqnarray}
    \item This operator is Baxterized using Method 3 of Sec.\ \ref{sec:Baxterization}, as it satisfies
    \begin{eqnarray}
         \check{Y}_{ij}^3=\check{Y}_{ij}^2+\alpha\beta~\check{Y}_{ij}-\alpha\beta~h_ih_j.
    \end{eqnarray}
   \item  Comparing this with the general form of the cubic constraint in \eqref{eq:cubic-minimal-poly}, results in the set of coupled polynomial equations for $\lambda_1$, $\lambda_2$ and $\lambda_3$ :
   \begin{eqnarray}
   \lambda_1+\lambda_2+\lambda_3 &=&1\nonumber\\
   \lambda_1\lambda_2+\lambda_2\lambda_3+\lambda_1\lambda_3&=&-\alpha\beta\nonumber\\
   \lambda_1\lambda_2\lambda_3&=&-\alpha\beta.
\end{eqnarray}
This yields the simplified equation
\begin{equation}\label{eq:constraint-eq}
(\lambda_2+\lambda_3)(\lambda_2-1)=0. 
\end{equation}
We do not use the solution $\lambda_2=-\lambda_3$, as it reduces the constraint in \eqref{eq:rmatrix-constraint} to 0. This makes the Hamiltonian in \eqref{eq:h-3-eigenvalues}, ill defined. We are then left with the choice
\begin{eqnarray}
    \{\lambda_2=1, \lambda_1=-\lambda_3=\pm \sqrt{\alpha\beta}\}.
\end{eqnarray}
%Other choices do not give a regular solution as they make the term
%\begin{equation} %\lambda_1+\lambda_2+\lambda_3+\lambda_1\lambda_3\lambda_2^{-1}= 0,
%\end{equation}
%leading to non-regular $\check{R}$-matrices [See Sec.\ \ref{subsec:3-eigenvalues}]. 

Using this, the algebraic form of the Hamiltonian is obtained from \eqref{eq:h-3-eigenvalues} by replacing $\check{Y}$ and $\check{Y}^{-\iota}$ with their algebraic expressions,
\begin{eqnarray}
         H=\frac{c}{1-\alpha\beta}~\sum\limits_j \left[\prod_{k=1}^{j-1} h_k\right]~\left[\check{Y}_{j,j+1}+\alpha\beta~ \check{Y}_{j,j+1}^{-\iota}-(1-\alpha\beta)~h_jh_{j+1}\right]~\left[\prod_{k=j+2}^{L} h_k\right]. \nonumber \\
     \end{eqnarray}
     \item The representation independent $R$-matrix is given by
\begin{equation} \label{eq:R-matrix-H14}
    \check{R}^{(1,4)}_{ij}(u)=(1-\alpha\beta)e^{-cu}~h_ih_j-(e^{-cu}-1)\check{Y}^{(1,4)}_{ij}-\alpha\beta~e^{-cu}(e^{-cu}-1)\check{Y}_{ij}^{(1,4)-\iota}.
\end{equation}
\item The $4\times 4$ version of the $R$-matrix is
\begin{equation}
\check{R}^{(1,4)}(u)=
\begin{pmatrix}
 (1-\alpha  \beta ) e^{-c u} & 0 & 0 & \alpha(1- e^{-2 c u}) \\
 0 & 1-\alpha  \beta  e^{-2 c u} & 0 & 0 \\
 0 & 0 & 1-\alpha  \beta  e^{-2 c u} & 0 \\
 \beta(1-e^{-2 c u}) & 0 & 0 & (1-\alpha  \beta ) e^{-c u} \\
\end{pmatrix}.
\end{equation}
\item In the $\mathbb{C}^2$ representation the Hamiltonian density derived from \eqref{eq:h-3-eigenvalues} is 
\begin{eqnarray}
    \mathfrak{h}_{(1,4)} 
%&=& \frac{(3\alpha\beta-1)}{2}\mathbb{1}+\frac{(\alpha+\beta)}{2}(\sigma^x_j\sigma^x_{j+1}-\sigma^y_j\sigma^y_{j+1}) \nonumber \\ &-& \frac{(1+\alpha\beta)}{2}\sigma^z_j\sigma^z_{j+1}+\frac{\mathrm{i}(\alpha-\beta)}{2}(\sigma^x_j\sigma^y_{j+1}+\sigma^y_j\sigma^x_{j+1}).
= 2\alpha~\sigma_j^+\sigma_{j+1}^+ + 2\beta~\sigma_j^-\sigma_{j+1}^- - \frac{(1+\alpha\beta)}{2}\sigma^z_j\sigma^z_{j+1}.
\end{eqnarray}
%&=& 
%\begin{pmatrix}
 %\alpha  \beta -1 & 0 & 0 & 2 \alpha  \\
 %0 & 2 \alpha  \beta  & 0 & 0 \\
 %0 & 0 & 2 \alpha  \beta  & 0 \\
 %2 \beta  & 0 & 0 & \alpha  \beta -1 \\
%\end{pmatrix}. \nonumber\\ \nonumber\\
It falls into the special case of the XYZ-type models.
%\begin{equation} 
%\mathfrak{h}_{(1,4)}=\kappa\left[Q\otimes Q\right]\begin{pmatrix}
 %a_1 & 0 & 0 & d_1 \\
 %0 & s_1 & c_1 & 0 \\
 %0 & c_1 & s_1 & 0 \\
 %d_2 & 0 & 0 & a_1 \\
%\end{pmatrix}\left[Q\otimes Q\right]^{-1}, 
%\end{equation}
%{\it via} the following local gauge transformations
%$$\left\{g_2\to 0,g_3\to 0,a_1\to \frac{\alpha  \beta -1}{\kappa },s_1\to \frac{2 \alpha  \beta }{\kappa },c_1\to 0,d_1\to \frac{2 \alpha  g_4^2}{g_1^2 \kappa },d_2\to \frac{2 \beta  g_1^2}{g_4^2 \kappa }\right\}.$$
\end{itemize}
\begin{remark}
Consider the constant braid operator 
\begin{eqnarray}
         \check{Y}_{ij}= \alpha~q_iq_j+\frac{1}{\alpha}~q^\dag_iq^\dag_j+b_i f_j+f_ib_j.
    \end{eqnarray}
This is one of the subclasses of $H1,4$\footnote{Another subclass obtained by taking $\alpha=\beta$ was Baxterized in \cite{Arnaudon_2003}.}. In this case, we will obtain a Baxterized solution under the Baxterization method 1\eqref{eq:Baxterization-condition-1}. 

%Thus, the local Hamiltonian expressed in \eqref{eq:h-method-1} takes the form in $\mathbb{C}^2$ as follows
%\begin{equation}
%\begin{pmatrix}
 %0 & 0 & 0 & \alpha  \\
 %0 & 1 & 0 & 0 \\
 %0 & 0 & 1 & 0 \\
 %\frac{1}{\alpha } & 0 & 0 & 0 \\
%\end{pmatrix}.
%\end{equation}
\end{remark}
%%%%%%%%%%%%%%%%%%%%%%%%%%%%%%%%%%%%%
\subsubsection{H3,1 class:}
\label{subsec:H31}
%%%%%%%%%%%%%%%%%%%%%%%%%%%%%%%%%%%%%
\begin{itemize}
    \item The $4\times 4$ braid operator is given by
    \begin{eqnarray}
     \check{Y}^{(3,1)} =    \begin{pmatrix}
 \alpha  & 0 & 0 & 0 \\
 0 & 0 & \gamma  & 0 \\
 0 & \delta  & 0 & 0 \\
 0 & 0 & 0 & \beta  \\
    \end{pmatrix}.
    \end{eqnarray}
    \item The corresponding SUSY expression is
    \begin{eqnarray}
         \check{Y}_{ij}^{(3,1)}=\alpha~b_i b_j+\beta~f_if_j+\gamma~q_iq^\dag_j+\delta~q^\dag_iq_j.
    \end{eqnarray}
    \item It can be Baxterized using Method 4 of Sec.\ \ref{sec:Baxterization}, as the braid operator satisfies
    \begin{eqnarray}
         \check{Y}^4=(\alpha+\beta)\check{Y}^3-(\alpha\beta-\gamma\delta)\check{Y}^2-\gamma\delta(\alpha+\beta)\check{Y}+\alpha\beta\gamma\delta~(h\otimes h).
    \end{eqnarray}
    Comparing this with \eqref{eq:quartic-relation} leads to the values of the $\lambda$'s :
    $$ \left\{\lambda_1, \lambda_2, \lambda_3, \lambda_4 \right\} = \{\alpha, \beta, \pm \sqrt{\gamma\delta}\}.  $$
    The integrable Hamiltonian corresponding to this solution is given in \eqref{eq:h-method-4}. This Hamiltonian is not well defined when one of the pairs $\{\lambda_1, \lambda_2\}$, $\{\lambda_2, \lambda_3\}$, $\{\lambda_3, \lambda_4\}$ assumes values $\pm \sqrt{\gamma\delta}$. However, the $R$-matrix for this solution given in \eqref{eq:rmatrix-method-4} satisfies the braided QYBE \eqref{eq:braided-YBE} in the entire parameter space only for this choice. To avoid this clash we have to choose $\alpha=\beta$ [scaled to 1]. Then the braid operator becomes 
     \begin{eqnarray}
         \check{Y}_{ij}^{(3,1)}&=&b_i b_j+f_if_j+\gamma~q_iq^\dag_j+\delta~q^\dag_iq_j,\\
         \check{Y}_{ij}^{(3,1)-\iota}&=&b_i b_j+f_if_j+\frac{1}{\delta}~q_iq^\dag_j+\frac{1}{\gamma}~q^\dag_iq_j.
    \end{eqnarray}
    This satisfies a cubic constraint 
    \begin{equation}
        \check{Y}_{ij}^3=\check{Y}_{ij}^2+\gamma \delta~ \check{Y}_{ij}-\gamma \delta~h_ih_j.
    \end{equation}
    Comparing this with the cubic constraint in \eqref{eq:cubic-minimal-poly} we find that
     $$\{\lambda_1, \lambda_2, \lambda_3\}=\{1,\pm \sqrt{\gamma \delta}\}.$$ 
     Among the many possibilities only the choices
     $$\{\lambda_1=-\lambda_3=\sqrt{\gamma\delta},~\lambda_2=1\}~\textrm{and}~\{\lambda_1=-\lambda_3=-\sqrt{\gamma\delta},~\lambda_2=1\},$$ result in well-defined Hamiltonians [See Sec.\ref{subsec:3-eigenvalues}]. Thus we can use the expression of the Hamiltonian given in \eqref{eq:h-3-eigenvalues} to find the algebraic expression for the integrable Hamiltonian as 
     \begin{eqnarray}
         H=\frac{c}{1-\gamma\delta}~\sum\limits_j \left[\prod_{k=1}^{j-1} h_k\right]~\left[\check{Y}_{j,j+1}+\gamma\delta~ \check{Y}_{j,j+1}^{-\iota}-(1-\gamma\delta)~h_jh_{j+1}\right]~\left[\prod_{k=j+2}^{L} h_k\right]. \nonumber \\
     \end{eqnarray}
     \item The algebraic form of the $R$-matrix is
\begin{equation} \label{eq:R-matrix-H31}
    \check{R}^{(3,1)}_{ij}(u)=(1-\gamma\delta)e^{-cu}~h_ih_j-(e^{-cu}-1)\check{Y}^{(3,1)}_{ij}-\gamma\delta~e^{-cu}(e^{-cu}-1)\check{Y}_{ij}^{(3,1)-\iota}.
\end{equation}
\item Choosing the $\mathbb{C}^2$ representation, the $R$-matrix becomes
\begin{equation}
\check{R}^{(3,1)}(u)=
\begin{pmatrix}
 1-\gamma  \delta  e^{-2 c u} & 0 & 0 & 0 \\
 0 & (1-\gamma  \delta ) e^{-c u} & \gamma(1-e^{-2 c u}) & 0 \\
 0 & \delta(1-e^{-2 c u}) & (1-\gamma  \delta ) e^{-c u} & 0 \\
 0 & 0 & 0 & 1-\gamma  \delta  e^{-2 c u} \\
\end{pmatrix}.
\end{equation}
    \item Then the corresponding Hamiltonian density is given by
\begin{eqnarray}
 \mathfrak{h}_{(3,1)}
 %   &=& \frac{(3\gamma\delta-1)}{2}\mathbb{1}+\frac{(\gamma+\delta)}{2}(\sigma^x_j\sigma^x_{j+1}+\sigma^y_j\sigma^y_{j+1}) \\ \nonumber  & +& \frac{(1+\gamma\delta)}{2}\sigma^z_j\sigma^z_{j+1}+\frac{\mathrm{i}(\gamma-\delta)}{2}(\sigma^y_j\sigma^x_{j+1}-\sigma^x_j\sigma^y_{j+1}).
 = 2\gamma~\sigma_j^+\sigma_{j+1}^- + 2\delta~\sigma_j^-\sigma_{j+1}^+ + \frac{(1+\gamma\delta)}{2}\sigma^z_j\sigma^z_{j+1}.
\end{eqnarray}
This Hamiltonian density is equivalent with the $XXZ$ model. %\eqref{eq:xxz-model-1} via the following gauge transformations
%$$\left\{g_1\to 0,g_4\to 0,a_1\to \frac{2 \gamma  \delta }{\kappa },s_1\to \frac{\gamma  \delta -1}{\kappa },c_1\to \frac{2 \delta }{\kappa },s_2\to \frac{\gamma  \delta -1}{\kappa },c_2\to \frac{2 \gamma }{\kappa }\right\}.$$
%It is important to emphasize that while the Hamiltonian of $H2,1$ is categorized under the special case of \eqref{eq:xxz-model-1}, $H3,1$ is not equivalent with $H2,1$.
\end{itemize}
%&=& \begin{pmatrix}
 %2 \gamma  \delta  & 0 & 0 & 0 \\
 %0 & \gamma  \delta -1 & 2 \gamma  & 0 \\
 %0 & 2 \delta  & \gamma  \delta -1 & 0 \\
 %0 & 0 & 0 & 2 \gamma  \delta  \\
  %  \end{pmatrix},\nonumber\\ \nonumber\\

\begin{remark}
Consider the constant braid operator 
\begin{eqnarray}\label{eq:H31-subclass-method-1}
         \check{Y}_{ij}= b_i b_j+f_if_j+\gamma~q_iq^\dag_j+\frac{1}{\gamma}~q^\dag_iq_j.
    \end{eqnarray}
This is one of the subclasses of $H3,1$. This can be Baxterized using Method 1 of Sec.\ \ref{sec:Baxterization} as it satisfies \eqref{eq:Baxterization-condition-1}. 

%Thus, by using the expression given in \eqref{eq:h-method-1}, we get the local Hamiltonian in $\mathbb{C}^2$ as
%\begin{equation}
%\begin{pmatrix}
 %1 & 0 & 0 & 0 \\
 %0 & 0 & \gamma  & 0 \\
 %0 & \frac{1}{\gamma } & 0 & 0 \\
 %0 & 0 & 0 & 1 \\
%\end{pmatrix}.
%\end{equation}
\end{remark}

%%%%%%%%%%%%%%%%%%%%%%%%%%%%%%%%%%%%%%%%%%%%%%%%%%%%%%%%%%%%%%%%%%%%%%
\subsection{Non-diagonalizable solutions}
\label{sec:Models-nondiagonalizable}
%%%%%%%%%%%%%%%%%%%%%%%%%%%%%%%%%%%%%%%%%%%%%%%%%%%%%%%%%%%%%%%%%%%%%%
We will now examine the two invertible classes of Hietarinta's classification that are not diagonalizable. We use Method 5 of Sec.\ \ref{sec:Baxterization} to deal with these classes.

%%%%%%%%%%%%%%%%%%%%%%%%%%%%%%%%%%%
\subsubsection{H0,1 class:} 
\label{subsec:H01}
%%%%%%%%%%%%%%%%%%%%%%%%%%%%%%%%%%%
\begin{itemize}
    \item The $4\times 4$ representative of this class is given by
    \begin{eqnarray}
         \check{Y}^{(0,1)} = \begin{pmatrix}
 1 & 0 & 0 & 1 \\
 0 & 0 & -1 & 0 \\
 0 & -1 & 0 & 0 \\
 0 & 0 & 0 & 1 \\
    \end{pmatrix}.
    \end{eqnarray}
    \item Its SUSY expression is 
    \begin{eqnarray}
        \check{Y}_{ij}^{(0,1)}=b_ib_j+f_if_j-(q_iq^\dag_j+q^\dag_iq_j)+q_iq_j.
    \end{eqnarray}
    \item This is Baxterized by Method 5 of Sec.\ \ref{subsubsec:nilpotent-1} as the braid operator satisfies 
    \begin{eqnarray}\label{eq:square-Y01}
        \check{Y}_{ij}^2 =h_ih_j+ N_{ij},~~N_{ij}=2q_iq_j.
    \end{eqnarray}
    \item This leads to the following $R$-matrix in terms of SUSY algebra elements
    \begin{equation}\label{eq:R-matrix-H01}
    \check{R}^{(0,1)}_{ij}(u)= h_ih_{j}+cu \left(\check{Y}^{(0,1)}_{ij}-\frac{N_{ij}}{2}\right).
\end{equation}
    \item The algebraic expression for the integrable Hamiltonian then follows from the constant regular $R$-matrix \eqref{eq:nilpotent-R-1}
    \begin{eqnarray}
        H= c~\left[\prod_{k=1}^L h_k\right]~\sum\limits_j \left[\check{Y}_{j,j+1}-\frac{N_{j,j+1}}{2}\right].
    \end{eqnarray}
    \item In the $\mathbb{C}^2$ representation, the $R$-matrix takes the form
\begin{equation}
\check{R}^{(0,1)}(u)=
\begin{pmatrix}
 c u+1 & 0 & 0 & 0 \\
 0 & 1 & -c u & 0 \\
 0 & -c u & 1 & 0 \\
 0 & 0 & 0 & c u+1 \\
\end{pmatrix}.
\end{equation}
    \item In the $\mathbb{C}^2$ representation the local Hamiltonian density is 
    \begin{eqnarray}
 \mathfrak{h}_{(0,1)}
    &=& \frac{1}{2}(\mathbb{1}-\sigma^x_j\sigma^x_{j+1}-\sigma^y_j\sigma^y_{j+1}+\sigma^z_j\sigma^z_{j+1}).
\end{eqnarray}
%&=& \begin{pmatrix}
 %1 & 0 & 0 & 0 \\
 %0 & 0 & -1 & 0 \\
 %0 & -1 & 0 & 0 \\
 %0 & 0 & 0 & 1 \\
  %  \end{pmatrix} \nonumber\\ \nonumber\\
This maps to the usual Heisenberg $XXX$-spin chain by a unitary rotation.
%\begin{equation}
 %   \mathcal{U} =\prod\limits_{j\in\textrm{odd}} \sigma^z_j,
%\end{equation}
%up to an overall negative sign.
\end{itemize}

%%%%%%%%%%%%%%%%%%%%%%%%%%%%%%%%%%%%%%%
\subsubsection{H2,3 class:} 
\label{subsec:H23}
%%%%%%%%%%%%%%%%%%%%%%%%%%%%%%%%%%%%%%%
\begin{itemize}
    \item The $4\times 4$ braid operator is 
    \begin{eqnarray}
       \check{Y}^{(2,3)} =  \begin{pmatrix}
 1 & \alpha  & \beta  & \gamma  \\
 0 & 0 & 1 & \alpha  \\
 0 & 1 & 0 & \beta  \\
 0 & 0 & 0 & 1 \\
    \end{pmatrix}.
    \end{eqnarray}
    \item The SUSY version is given by
    \begin{eqnarray}
         \check{Y}_{ij}=(b_ib_j+f_if_j+q^\dag_iq_j+q_iq^\dag_j)+\alpha~(b_iq_j+q_if_j)+\beta~(q_ib_j+f_iq_j)+\gamma~q_iq_j.
    \end{eqnarray}
    \item This is Baxterized using Method 5 of Sec.\ \ref{subsubsec:nilpotent-2}, as in this case the nilpotent operator
    \begin{eqnarray}\label{eq:nilpotent23}
         N_{ij}=(\alpha+\beta)(h_iq_j+q_ih_j)+(\alpha^2+\beta^2+2\gamma)q_iq_j,
    \end{eqnarray}
    satisfies $N^3=0$.
    \item We are then lead to the following algebraic version of the $R$-matrix
    \begin{equation}\label{eq:R-matrix-H23}
    \check{R}^{(2,3)}_{ij}(u)=h_i h_j+ cu\left( \check{Y}^{(2,3)}_{ij}-\frac{1}{2} N_{ij}+\frac{1}{8} N_{ij}^2\right).
\end{equation}
    \item The integrable Hamiltonian is obtained from the regular $R$-matrix in \eqref{eq:nilpotent-R-2} :
    \begin{eqnarray}
        H= c~\left[\prod_{k=1}^L h_k\right]~\sum\limits_j \left[\check{Y}_{j,j+1}-\frac{N_{j,j+1}}{2}+\frac{N^2_{j,j+1}}{8}\right].
    \end{eqnarray}
    \item In the $\mathbb{C}^2$ representation, the $R$-matrix becomes
\begin{equation}
\check{R}^{(2,3)}(u)=
\begin{pmatrix}
 c u+1 & \frac{1}{2} c u (\alpha -\beta ) & \frac{1}{2} c u (\beta -\alpha ) & -\frac{1}{4} c u (\alpha -\beta )^2 \\
 0 & 1 & c u & \frac{1}{2} c u (\alpha -\beta ) \\
 0 & c u & 1 & \frac{1}{2} c u (\beta -\alpha ) \\
 0 & 0 & 0 & c u+1 \\
\end{pmatrix}.
\end{equation}
    \item In the $\mathbb{C}^2$ representation, the Hamiltonian density becomes 
   % \begin{eqnarray}
    %     \mathfrak{h}_{(2,3)}
    %&=& \frac{1}{2}(\mathbb{1}+\sigma^z_j\sigma^z_{j+1})+\frac{(8-(\alpha-\beta)^2)}{16}\sigma^x_j\sigma^x_{j+1}+\frac{(8+(\alpha-\beta)^2)}{16}\sigma^y_j\sigma^y_{j+1}\nonumber\\&-&\frac{\mathrm{i}(\alpha-\beta)^2}{16}(\sigma^x_j\sigma^y_{j+1}+\sigma^y_j\sigma^x_{j+1})+\frac{(\alpha-\beta)}{2}(\sigma^z_j\sigma^+_{j+1}-\sigma^+_j\sigma^z_{j+1}).
    %\end{eqnarray}
    \begin{equation}
\mathfrak{h}_{(2,3)}= P_{j,j+1}-\frac{(\alpha-\beta)^2}{4}~\sigma^+_j\sigma^+_{j+1}+\frac{(\alpha-\beta)}{2}(\sigma^z_j\sigma^+_{j+1}-\sigma^+_j\sigma^z_{j+1}).
\end{equation} 
    This Hamiltonian density does not fall into any of the known integrable models. However, this represents the special case of the Hamiltonian density in \eqref{eq:hdensity-H13} obtained by setting $\alpha=1$ and $\beta=-\gamma$.
\end{itemize}
%&=&\begin{pmatrix}
 %1 & \frac{\alpha -\beta }{2} & \frac{\beta -\alpha }{2} & -\frac{1}{4} (\alpha -\beta )^2 \\
% 0 & 0 & 1 & \frac{\alpha -\beta }{2} \\
% 0 & 1 & 0 & \frac{\beta -\alpha }{2} \\
% 0 & 0 & 0 & 1 \\
 %   \end{pmatrix} \nonumber\\ \nonumber\\

\begin{remark}
    Setting $\alpha=-\beta$ in \eqref{eq:nilpotent23} changes the nilpotency order of $N$ from $r=3$ to $r=2$. For this choice of parameters we will then need to use the Baxterization method of Sec.\ \ref{subsubsec:nilpotent-1}. This leads to a new set of $R$-matrices that are inequivalent to those obtained from the $H0,1$ class. The Hamiltonians in this case are 
   \begin{equation}
        H= c~\left[\prod_{k=1}^L h_k\right]~\sum\limits_j \left[\check{Y}'_{j,j+1}-\frac{\tilde{N}_{j,j+1}}{2}\right]
   \end{equation}
   Where \begin{equation}
    \check{Y}'_{ij}=(b_ib_j+f_if_j+q^\dag_iq_j+q_iq^\dag_j)-\beta~(b_iq_j+q_if_j-q_ib_j-f_iq_j)+\gamma~q_iq_j.
   \end{equation}
   \begin{equation}
       \tilde{N}=2 (\beta^2+\gamma)q_iq_j
   \end{equation}
   % In $\mathbb{C}^2$, the local term of the Hamiltonian is
    %\begin{equation}
     %   \begin{pmatrix}
 %1 & -\beta  & \beta  & -\beta ^2 \\
 %0 & 0 & 1 & -\beta  \\
 %0 & 1 & 0 & \beta  \\
 %0 & 0 & 0 & 1 \\
  %      \end{pmatrix}.
  %  \end{equation} 
\end{remark}

%%%%%%%%%%%%%%%%%%%%%%%%%%%%%%%%%%%%%%%%%%%%%%%%%%%%%%%%%%%%%%%
\subsection{Summary of the spin $\frac{1}{2}$ Hamiltonians}
\label{sec:summary}
%%%%%%%%%%%%%%%%%%%%%%%%%%%%%%%%%%%%%%%%%%%%%%%%%%%%%%%%%%%%%%%
In Secs. \ref{sec:Models-diagonalizable} and \ref{sec:Models-nondiagonalizable} we have constructed 7 Hamiltonian densities in the $\mathbb{C}^2$ representation that can be mapped to known integrable spin $\frac{1}{2}$ models. These are summarized in Table \ref{tab:old-h}. We have also discovered two new integrable spin $\frac{1}{2}$ systems with nearest-neighbor interactions as summarized in Table \ref{tab:new-h}. The higher spin analogs of these integrable models will be considered in Sec.\ \ref{sec:higher-spin}. 

\begin{longtable}{ |c|c|c| }
 \hline 
 Class & $4 \times 4$ $\mathfrak{h}$ & $n$-vertex model\\
 \hline \hline
 $H0,1$ & $\frac{1}{2}\left[\mathbb{1}-\sigma^x_j\sigma^x_{j+1}-\sigma^y_j\sigma^y_{j+1}+\sigma^z_j\sigma^z_{j+1}\right]$ & 6, [$XXX$]\\
    \hline
 $H0,2$ & $\mathbb{1}+\mathrm{i}~\sigma^y_j\sigma^x_{j+1}$ & 8 \\ 
    \hline
 $H1,1$ & $\beta^2~ \sigma^x_j\sigma^x_{j+1}+\gamma^2~\sigma^y_j\sigma^y_{j+1}+\beta\gamma~\left[\sigma^z_j+\sigma^z_{j+1}\right]$ & 8 \\
 \hline
 $H1,2$ & \parbox{0.4\textwidth}{\begin{eqnarray*}
    &&  \alpha~\sigma_j^+\sigma_{j+1}^+ + \beta~\sigma_j^+\sigma_{j+1}^- + \gamma~\sigma^-_j\sigma^+_{j+1} \\ & + &  \frac{1}{2}\left[\beta~\sigma^z_j+\gamma~\sigma^z_{i+1}\right]
 \end{eqnarray*}} & 7\\
 \hline
 $H1,4$ & \parbox{0.5\textwidth}{\begin{eqnarray*}
    &&  2\alpha~\sigma_j^+\sigma_{j+1}^+ + 2\beta~\sigma_j^-\sigma_{j+1}^- - \frac{(1+\alpha\beta)}{2}\sigma^z_j\sigma^z_{j+1}
 \end{eqnarray*}} & 8 \\
 \hline
 $H2,1$ & \parbox{0.5\textwidth}{\begin{eqnarray*}
    && \alpha\beta~\sigma_j^+\sigma_{j+1}^- + \alpha\gamma~\sigma_j^-\sigma_{j+1}^+ + \frac{(\alpha^2+\beta\gamma)}{4}\sigma^z_j\sigma^z_{j+1}
 \end{eqnarray*}} & 6, [$XXZ$] \\
 \hline
 $H3,1$ & \parbox{0.5\textwidth}{\begin{eqnarray*}
    && 2\gamma~\sigma_j^+\sigma_{j+1}^- + 2\delta~\sigma_j^-\sigma_{j+1}^+ + \frac{(1+\gamma\delta)}{2}\sigma^z_j\sigma^z_{j+1}
 \end{eqnarray*}} & 6, [$XXZ$] \\
 \hline
 \caption{Closed chain, spin $\frac{1}{2}$ Hamiltonian densities for 7 equivalence classes. They can be mapped to known models.} 
 %Note that the Hamiltonian densities are to be identified with the $\check{W}$ operator in \eqref{eq:hW}. Thus they need not themselves be solutions of the braided QYBE.
 \label{tab:old-h}
 \end{longtable}

\begin{longtable}{ |c|c| }
 \hline 
 Class & New $4 \times 4$ $\mathfrak{h}$ \\
 \hline \hline
 $H1,3$ & \parbox{0.6\textwidth}{\begin{eqnarray*}
    && \alpha^2~P_{j,j+1}+\beta\gamma~\sigma_j^+\sigma_{j+1}^++\frac{\alpha(\beta-\gamma)}{2}\left[\sigma^+_j\sigma^z_{j+1}-\sigma^z_j\sigma^+_{j+1}\right]
 \end{eqnarray*}}\\
    \hline
 $H2,2$ & \parbox{0.5\textwidth}{\begin{eqnarray*}
    && \alpha\beta~\sigma^+_j\sigma^-_{j+1} + \alpha\gamma~\sigma^-_j\sigma^+_{j+1} + \frac{\alpha^2}{4}\sigma^z_j+\frac{\beta\gamma}{2}\sigma^z_{j+1}
 \end{eqnarray*}}\\
    \hline
 \caption{The new Hamiltonian densities in the $\mathbb{C}^2$ representation on a closed chain. $P$ is the standard permutation on $\mathbb{C}^2\otimes\mathbb{C}^2$. The Hamiltonian density corresponding to the $H2,3$ class is a special case of the $H1,3$ class, as explained in Sec. \ref{subsec:H23}, and so is not included in this list. }
    \label{tab:new-h}
\end{longtable}

Some important properties are:
\begin{enumerate}
    \item We find that out of ten Hamiltonians considered, only two, $H0,1$ and $H1,1$ are hermitian, while the remaining eight correspond to non-hermitian Hamiltonians. 
    \item The Hamiltonian corresponding to the $H1,3$ class is a non-hermitian deformation of the permutation model or $XXX$-spin chain. However, this Hamiltonian is non-diagonalizable, in contrast to the standard $XXX$-spin chain that is diagonalizable. This implies that the two Hamiltonians are inequivalent.
    \item Using a Jordan-Wigner transform, the models corresponding to the $H0,2$, $H1,1$, $H1,2$ and $H2,2$ classes, are free fermion systems. It should be noted that, though free fermion systems can be solved exactly without the need of sophisticated techniques like the algebraic Bethe ansatz, the fact that they can be obtained from a $R$-matrix is not immediate or obvious. The existence of a $R$-matrix for such systems can still reveal other interesting features. For instance, consider the critical Ising chain, another well-known free fermion system. Its transfer matrix can be constructed using a $R$-matrix and recently, this was shown to contain the non-invertible Kramers-Wannier duality symmetry \cite{Sinha_2025}.
  \item It is also important to highlight that despite identifying two constant braid operators corresponding to the Hietarinta's classes $H0,1$, $H2,3$, as non-diagonalizable, we are still able to construct diagonalizable Hamiltonian densities in each case. 
    \item We compared all the Hamiltonian densities with the $4 \times 4$ Hamiltonians presented in subsections (6.1-6.7) of \cite{de2019classifying}, with two additional models given in Eqns. 66 and 69 on page 13 of \cite{deLeeuw2024}, with the differential approach of \cite{vieira2018solving}, and with earlier works on classifications of spin $\frac{1}{2}$ integrable systems \cite{krichever1981baxter,Sogo}. We find that the Hamiltonians shown in Table \ref{tab:new-h} are not equivalent to any of those presented in these references. Among the equivalences we checked, we also included Drinfeld twists.
    %\item Furthermore, our analysis additionally indicates that the entire H1,3 class cannot be related, via a Drinfeld twist, to any of the known spin-$\frac{1}{2}$ Hamiltonians documented in the work of Marius et al., highlighting the fact that it constitutes a genuinely distinct integrable model.
\end{enumerate}

\section{Higher spin integrable systems}
\label{sec:higher-spin}
%%%%%%%%%%%%%%%%%%%%%%%%%%%%%%%%%%%%%%%%%%%%%%%%%%%%
Higher $d$-dimensional representations of the supercharges lead to spin $\frac{d-1}{2}$ integrable models. For each $d$, inequivalent representations of the $\mathcal{N}=2$ SUSY algebra, correspond to inequivalent Hamiltonian densities. So, our first task is to classify them. These representations depend on the binary [$\mathbb{Z}_2$-grading] partitions of $d$. If they are unequal, the SUSY Hamiltonian is a non-trivial projector and we obtain almost local models, as defined in Sec. \ref{sec:preliminaries}. On the other hand, when $d$ is partitioned into two subspaces of equal dimension, then the SUSY Hamiltonian is just the $d\times d$ identity operator, making the Hamiltonian local. These two cases sufficiently describe the different types of higher spin chains that can be obtained from the SUSY Baxterization technique. We illustrate both these cases with the Hamiltonians corresponding to the permutation operator. Note that the permutation operator forms another equivalence class in the classification of the $4\times 4$ constant braid solutions and was not considered in Sec. \ref{sec:hamiltonians}. These examples indicate the structure of the higher spin Hamiltonians for all the other Hietarinta equivalence classes. All of this is worked out for the $d=3$ [spin 1] and $d=4$ [spin $\frac{3}{2}$] cases. The choices for $d$ are made without any loss of generality.

%%%%%%%%%%%%%%%%%%%%%%%%%%%%%%%%%%%%%%%%%%%%%%%%%%%%%%%%%%%%%%%%%%%%%
\subsection{Classification of the inequivalent representations of the supercharges}
\label{subsec:classification-q-d}
%%%%%%%%%%%%%%%%%%%%%%%%%%%%%%%%%%%%%%%%%%%%%%%%%%%%%%%%%%%%%%%%%%%%%
Supercharges are nilpotent and thus, non-diagonalizable operators. They can be written in the Jordan normal form \cite{axler2024linear,horn2012matrix}, reducing their classification in a given dimension $d$, into finding all the inequivalent Jordan normal forms. This normal form is a direct sum of Jordan blocks, each with eigenvalue 0. The possible Jordan blocks for constructing $q$ are 
\begin{equation}
    J_1=\sigma^+=\begin{pmatrix}
        0 & 1 \\
        0 & 0
    \end{pmatrix}~;~~ J_0 =0.
\end{equation}
The rank of the matrix equals the number of $J_1$ blocks. 
The inequivalent representations are classified by the number of $J_1$ and $J_0$ blocks. Thus, there is one representation for every possible rank. The rank can be counted as follows. Let there be $m$, $J_1$ blocks in a given representation. Then, for a given $d$, the possible ranks are given by 
$$1\leq m \leq \left\lfloor \frac{d}{2}\right\rfloor.$$
The upper limit also coincides with the maximum possible rank of the supercharge $q$. The remaining positions in the supercharge will be filled with $d-2m$ number of the one dimensional $J_0$ blocks.

\begin{remark}
The Jordan normal form consists of a matrix with non-zero entries [mostly 1] on the upper-diagonal [a diagonal off the main diagonal]. However, this does not make the $\mathbb{Z}_2$ grading structure of the supercharge transparent. Henceforth, the supercharge in the Jordan normal form will be denoted $q_J$, distinguishing it from the supercharge $q$, which makes the $\mathbb{Z}_2$ grading structure apparent. The two forms of the supercharges are connected by a similarity transformation, $S$. 
\end{remark}

We will now illustrate the construction of the inequivalent supercharges through explicit examples. The cases $d=2,3,4$ are worked out. This is followed by the expression for the supercharges in a general $d$.

\begin{itemize}
    \item For $d=2$, the $\mathbb{Z}_2$ partition is $1+1$ and so the only possible rank is 1. The lone supercharge is given by
$$ q = q_J =\begin{pmatrix}
    . &  1 \\ .  & .
\end{pmatrix}  $$
This coincides with the $\mathbb{C}^2$ SUSY representation in \eqref{eq:susy-representation-C2}. See also Remark \ref{rem:nilpotent-qs-c2} in this regard.
     \item When $d=3$, the only allowed partition is $2+1$. The two forms of the corresponding rank 1 supercharge and the similarity transform between them are given by  
\begin{eqnarray}\label{eq:q-d-3}
 q_J = \begin{pmatrix}
        . & 1 & . \\ . & . & . \\ . & . & .
    \end{pmatrix} ~;~  q = \frac{1}{\sqrt{2}}\begin{pmatrix}
        . & 1 & 1 \\ . & . & . \\ . & . & .
    \end{pmatrix}~;~ S=\begin{pmatrix}
 1 & . & . \\
 . & \sqrt{2} & -1 \\
 . & . & 1 \\
    \end{pmatrix}.
\end{eqnarray}
\item For the $d=4$ case, we have two partitions possible, $3+1$ and $2+2$. The corresponding rank 1 and rank 2 supercharges and the similarity transform $S$ between their different forms are given by,
\begin{eqnarray}
    q_J = \begin{pmatrix}
    . & 1 & . & . \\ . & . & . & . \\ . & . & . & . \\ . & . & . & .
\end{pmatrix}~;~ q=\frac{1}{\sqrt{3}}\begin{pmatrix}
        . & 1 & 1 & 1 \\ . & . & . & . \\ . & . & . & . \\ . & . & . & . 
    \end{pmatrix}~;~S=\begin{pmatrix}
 1 & . & . & . \\
 . & \sqrt{3} & -1 & -1 \\
 . & . & . & 1 \\
 . & . & 1 & . \\
    \end{pmatrix},
\end{eqnarray}
and 
\begin{eqnarray}\label{eq:2+2supercharge}
    q_J=\begin{pmatrix}
    . & 1 & . & . \\ . & . & . & . \\ . & . & . & 1 \\ . & . & . & .
\end{pmatrix}~;~q=\begin{pmatrix}
        . & . & 1 & . \\ . & . & . & 1 \\ . & . & . & . \\ . & . & . & .
    \end{pmatrix}~;~S=\begin{pmatrix}
 . & . & 1 & . \\
 1 & . & . & . \\
 . & . & . & 1 \\
 . & 1 & . & . \\
    \end{pmatrix},
\end{eqnarray}
respectively.
\end{itemize}
The above examples suggest the general algorithm to construct a supercharge for a given rank. So, in an arbitrary $d$-dimensional representation, the rank $m<d$ supercharge corresponds to a $\mathbb{Z}_2$ graded partition $(d-m)+m$. Assuming that $m<d-m$, this supercharge is given by
\begin{eqnarray}
    q = \begin{pmatrix}
        0_{m\times m} & A \\ 0_{(d-m)\times m} & 0_{(d-m) \times (d-m)}
    \end{pmatrix} ~;~ A = \begin{pmatrix}
        \mathbb{1}_{(m-1)\times (m-1)} & 0_{(m-1)\times (d-2m +1) } \\ 0_{1\times (m-1)} &  \left(\frac{1}{\sqrt{d-2m+1}}\right)_{1\times (d-2m+1)}
    \end{pmatrix}
\end{eqnarray}
Using this method we can construct the $\lfloor\frac{d}{2}\rfloor$ inequivalent supercharges for a given $d$. Thus, in an arbitrary $d$, we obtain the same number of inequivalent spin $\frac{d-1}{2}$ chains for each Hietarinta class. 

From the structure of the different rank supercharges for a given $d$, it is clear that we can group them together into two types: unequal or equal partitions of $d$. For the former case, the corresponding SUSY Hamiltonian $h$, is a non-trivial projector, whereas for the latter it reduces to the $d\times d$ identity operator. Thus we see that for representations with unequal partitions, the global string operators in each term of the Hamiltonian remain, making the entire model almost local as defined in Sec. \ref{sec:preliminaries}. On the other hand for representations with equal partitions, these string operators become the identity, making the Hamiltonian local. Unequal partitions arise for odd and even $d$, whereas equal partitions are only obtained when $d$ is even. We will now consider examples of Hamiltonians for the two cases.

%%%%%%%%%%%%%%%%%%%%%%%%%%%%%%%%%%%%%%%%%%%%%%%%%%%%%%%%%%%%%%%%%%%%%
\subsection{Higher spin Heisenberg-like systems}
\label{subsec:higher-spin-permutation-models}
%%%%%%%%%%%%%%%%%%%%%%%%%%%%%%%%%%%%%%%%%%%%%%%%%%%%%%%%%%%%%%%%%%%%%
The $4\times 4$ expression for the permutation operator $P$, suggests that its SUSY realization is given by,
    \begin{eqnarray}\label{eq:perm-SUSY}
        P_{ij} = b_ib_j + f_if_j + q_iq_j^\dag + q_i^\dag q_j. 
    \end{eqnarray}
    Using the relations obeyed by the SUSY generators, \eqref{eq:susy-relations}, it is easy to verify that this permutation operator satisfies
    \begin{eqnarray}
        & P_{ij}P_{jk}P_{ij} = P_{jk}P_{ij}P_{jk} = P_{ik}h_j & \nonumber \\
        & P_{ij}^2  = h_ih_j.&
    \end{eqnarray}
    Here $h_j$ is the SUSY Hamiltonian. Note that in the $\mathbb{C}^2$ representation of the SUSY generators, \eqref{eq:susy-representation-C2}, these relations reduce to the usual permutation algebra, as the SUSY Hamiltonian is just the identity operator. We can Baxterize the permutation operator using Method 1 of Sec. \ref{subsec:method1-Baxterization}, resulting in the $R$-matrix
    \begin{eqnarray}
        \check{R}_{ij}(u) = h_ih_j + cu~P_{ij}~;~c,u\in\mathbb{C}.
    \end{eqnarray}
    The algebraic expression for the associated Hamiltonian yields the Heisenberg-like model given by,
    \begin{eqnarray}
        H = c\sum\limits_{j=1}^L~\left[\prod\limits_{k=1}^{j-1}~h_k \right]P_{j,j+1}\left[\prod\limits_{k=j+2}^{L}~h_k \right].
    \end{eqnarray} 
    The constant $c$ can be taken to be negative in order to describe a ferromagnetic system. The strings of $h$'s make this Hamiltonian almost local. We will consider this model in the spin 1 [unequal partition of $d=3$] and the spin $\frac{3}{2}$ [equal partition of $d=4$] cases.

%%%%%%%%%%%%%%%%%%%%%%%%%%%%%%%%%%%%%%%%%%%%%%%%%%%%%%%%%%%%%%%%
\subsubsection*{Spin 1 Heisenberg-like chain :}
\label{subsubsec:spin1-heisenberg-chain}
%%%%%%%%%%%%%%%%%%%%%%%%%%%%%%%%%%%%%%%%%%%%%%%%%%%%%%%%%%%%%%%%%%   
In the $\mathbb{C}^3$ representation, the supercharges are $q$, given by \eqref{eq:q-d-3}, and its adjoint. They swap the two $\mathbb{C}^3$ vectors,
\begin{eqnarray}\label{eq:c3-supercharge-action}
        \ket{u}\equiv\begin{pmatrix}
            1 \\ 0 \\ 0
        \end{pmatrix}~;~\ket{d}\equiv\frac{1}{\sqrt{2}}\begin{pmatrix}
            0 \\ 1 \\ 1
        \end{pmatrix}.
    \end{eqnarray}
These vectors span a two dimensional subspace of $\mathbb{C}^3$, resembling an effective spin $\frac{1}{2}$ system. For this reason we denote them $\ket{u}$($\ket{d}$) or more colloquially, the {\it up}({\it down}) states. Thus, this supercharge $\mathbb{Z}_2$ grades $\mathbb{C}^3$ into two subspaces of unequal dimensions. This implies that the Hamiltonian built using them will be almost local due to the presence of the string operators.

Using \eqref{eq:c3-supercharge-action}, we can deduce the action of the permutation operator, \eqref{eq:perm-SUSY}. We find that it is not the full permutation operator on $\mathbb{C}^3\otimes\mathbb{C}^3$, but instead only exchanges the vectors in the subspace spanned by tensor products of the $\ket{u}$ and $\ket{d}$ states. This makes it non-invertible. The third vector $\frac{1}{\sqrt{2}}\begin{pmatrix}
0 & 1 & -1 \end{pmatrix}^T$, is annihilated by the supercharges, and thus also by the SUSY Hamiltonian and the permutation operator. Thus this Hamiltonian mimics an effective Heisenberg-like model on $\mathbb{C}^3$, with the local terms exchanging the two-level system spanned by the states $\ket{u}$ and $\ket{d}$. All the remaining $3^L-2^L$ states, are zero modes of this system. The system has two ferromagnetic ground states given by the product states $\bigotimes\limits_{k=1}^L~\ket{u}_k$ and $\bigotimes\limits_{k=1}^L~\ket{d}_k$. With positive $c$, they have lower energy than the large number of zero modes of this system. The effective magnetization operator for this system is given by 
        \begin{eqnarray}
            M = \sum\limits_{k=1}^L~\tilde{Z}_k~;~\tilde{Z}_k=\textrm{Diag}~\left(1,-1,-1\right).
        \end{eqnarray} This operator distinguishes the two ferromagnetic states. 

%%%%%%%%%%%%%%%%%%%%%%%%%%%%%%%%%%%%%%%%%%%%%%%%%%%%%%%%%%%%%%%%
\subsubsection*{Spin $\frac{3}{2}$ Heisenberg-like chain :}
\label{subsubsec:spin3/2-heisenberg-chain}
%%%%%%%%%%%%%%%%%%%%%%%%%%%%%%%%%%%%%%%%%%%%%%%%%%%%%%%%%%%%%%%%%%    
In this case the dimensions of the partitions are each equal to 2, implemented by the supercharge $q$ of \eqref{eq:2+2supercharge} and its adjoint. They have the following action on the canonical basis of $\mathbb{C}^4$:
\begin{eqnarray}\label{eq:c4-supercharge-action}
    \begin{pmatrix}
        1 \\ 0 \\ 0 \\ 0
    \end{pmatrix} \leftrightarrow \begin{pmatrix}
        0 \\ 0 \\ 1 \\ 0
    \end{pmatrix}~;~\begin{pmatrix}
        0 \\ 1 \\ 0 \\ 0
    \end{pmatrix} \leftrightarrow \begin{pmatrix}
        0 \\ 0 \\ 0 \\ 1
    \end{pmatrix}.
\end{eqnarray}
As this exchange action involves all the basis vectors of $\mathbb{C}^4$, the SUSY Hamiltonian $h$, is no longer a projector, but the identity operator on $\mathbb{C}^4$.  
Using this we deduce the action of the permutation operator \eqref{eq:perm-SUSY} on $\mathbb{C}^4\otimes\mathbb{C}^4$. The result is more conveniently expressed when we consider each $\mathbb{C}^4$ as $\mathbb{C}^2\otimes\mathbb{C}^2$. The action on two sites $i$, $j$ is shown in Figure \ref{fig:doubled-lattice}.
\begin{figure}[h]
    \centering
    \begin{tikzpicture}[scale=1]
        % lattice sites
        \foreach \x in {2,...,5} {
            \fill (\x,0) circle (2.5pt);
            %\node[below] at (\x,0) {\x};
        }
        % chain connection
        \draw[thick] (1,0) -- (6,0);
        % ellipse around sites 2 and 3
    \draw[red,thick] (2.5,0) ellipse (0.75cm and 0.35cm);
    \node[below] at (2.5,-0.4) {$i$};
     % ellipse around sites 4 and 5
    \draw[red,thick] (4.5,0) ellipse (0.75cm and 0.35cm);
    \node[below] at (4.5,-0.4) {$j$};
        \draw[->,thick] (4,0) -- ++(-0.1,0.8) node[left] {$j_I$};
        \draw[->,thick] (5,0) -- ++(0.1,0.8) node[right] {$j_{II}$};
\end{tikzpicture}
    \caption{Sites of the doubled lattice. Each of sites $i$ and $j$ denoted by the ellipses, carries a $\mathbb{C}^4$. Each site inside the ellipse carries a $\mathbb{C}^2$.  }
    \label{fig:doubled-lattice}
\end{figure}
It shows that the spin $\frac{3}{2}$ model can be reinterpreted as a spin $\frac{1}{2}$ Hamiltonian with a 4-site interaction. Thus each site of the original $L$-site chain now gets doubled. We will denote the two parts of this doubling as $I$ and $II$ respectively. Then the site index $j$, of the original chain gets the additional suffixes $j_I$ and $j_{II}$. On this doubled lattice, the permutation operator acts as 
\begin{eqnarray}
    P_{ij} = p_{i_I,j_I}\otimes\mathbb{1}_{i_{II}, j_{II}},
\end{eqnarray} 
where $P$($p$) is the permutation operator on $\mathbb{C}^{4(2)}\otimes\mathbb{C}^{4(2)}$. Thus the model resembles the spin $\frac{1}{2}$ Heisenberg model on the indices with suffix $I$ and the identity action on the indices with suffix $II$. Unlike the unequal partition case, this model has no zero modes.

%%%%%%%%%%%%%%%%%%%%%%%%%%%%%%%%%%%%%%%%%%%%%%%%%%%%%%%%%%%%%%%%%%%%%
\subsection{Higher spin Hamiltonians of the other  Hietarinta classes}
\label{subsec:higher-spin-chains-AllClasses}
%%%%%%%%%%%%%%%%%%%%%%%%%%%%%%%%%%%%%%%%%%%%%%%%%%%%%%%%%%%%%%%%%%%%%
The higher spin Hamiltonians for the permutation operator help us to write down the higher spin versions of the Hamiltonians corresponding to the other Hietarinta classes [Tables \ref{tab:old-h}, \ref{tab:new-h}], without much more work. For the spin 1 case [unequal partitions of $d=3$], the higher spin Hamiltonian density for each class is similar to those given by Tables \ref{tab:old-h}, \ref{tab:new-h}, but they now act on the effective spin $\frac{1}{2}$ states spanned by the $\ket{u}$ and $\ket{d}$ in \eqref{eq:c3-supercharge-action}. The remaining states of the total Hilbert space are zero modes, as was the case in the spin 1 Heisenberg-like model. For the spin $\frac{3}{2}$ case [equal partitions of $d=4$], the Hamiltonian can be rewritten as a 4-site interaction of spin $\frac{1}{2}$ systems [See Figure \ref{fig:doubled-lattice}]. The Hamiltonian in this case just becomes 
\begin{eqnarray}
    H = \sum\limits_{j=1}^L~\mathfrak{h}_{j_I, (j+1)_{I}}\otimes \mathbb{1}_{j_{II}, (j+1)_{II}},
\end{eqnarray}
where the spin $\frac{1}{2}$ Hamiltonian density is taken from those listed in Tables \ref{tab:old-h} and \ref{tab:new-h}. This system has no zero modes when compared to the unequal partition case.

The Hamiltonians for other values of $d$ can be analogously constructed. From the above higher spin examples, we expect two types of Hamiltonians : in the case of unequal partitions of the $\mathbb{C}^d$, the higher spin Hamiltonian will be effectively an almost local spin $\frac{1}{2}$ system with a lot of zero modes and in the case of equal partitions, we can rewrite the nearest-neighbor higher spin interaction equivalently as a multi-site spin $\frac{1}{2}$ system, with no zero modes.

%%%%%%%%%%%%%%%%%%%%%%%%%%%%%%%%%%%%%%%%%%%%%%%%%%%%
\section{Conclusion}
\label{sec:conclusion}
%%%%%%%%%%%%%%%%%%%%%%%%%%%%%%%%%%%%%%%%%%%%%%%%%%%%
We have demonstrated how $\mathcal{N}=2$-SUSY algebras can be used to systematically construct algebraic versions of $4\times 4$ constant Yang-Baxter solutions or braid operators. Depending on the representation chosen for the SUSY generators, these braid operators are either invertible or non-invertible. The former case appears only in Hilbert spaces with even dimensions that are $\mathbb{Z}_2$-graded into two equal parts. In the remaining cases, the braid operators are non-invertible, but in a controlled way. By the latter, we mean that the braid operators obey a modified version of known quotients of the braid group, like the Iwahori-Hecke algebra. This makes these operators suitable for Baxterization, which eventually leads to integrable models. In the non-invertible case, the Hamiltonian densities are non-local but again in a very controlled manner. We call this almost local. As a result, many of the results used in the study of integrable systems, like the construction of conserved quantities using the boost operator method, still hold for these systems. 

With the techniques developed in this work, we have shown two new nearest-neighbor spin $\frac{1}{2}$ systems. They are obtained from the $(1,3)$, and $(2,2)$ Hietarinta classes as shown in Table \ref{tab:new-h}. The remaining 7 classes produce spin $\frac{1}{2}$ chains that belong to either one of the 6-, 7- or 8-vertex models, as shown in Table \ref{tab:old-h}. We have also provided the algorithm to construct the inequivalent set of supercharges in higher dimensions, which can then be used to construct higher-spin integrable models. 

Some future directions are as follows:
\begin{enumerate}
    \item The methods developed in this work can be repeated for the algebraic solutions for the $4\times 4$ constant Yang-Baxter solutions found in \cite{MSPK-Hietarinta}. In this case, the algebras used include partition algebras \cite{martin1994temperley,halverson2004partitionalgebras,Padmanabhan_2020} and Clifford algebra solutions \cite{padmanabhan2024solving}. We expect to find local Hamiltonians in these cases that do not coincide or are equivalent to the ones found in this work. This would also positively demonstrate the hypothesis that a given braid operator could be Baxterized in multiple ways.
    \item As the spin chains we obtain are almost a sum of braid group generators, it would be interesting to study their spectral properties by suitably adapting the algebraic Bethe ansatz method. We obtain Hamiltonians from non-invertible $R$-matrices, and so we expect some of them to be non-diagonalizable as well. In such cases, the usual methods of finding the spectrum with the help of the algebraic Bethe ansatz fail. Nevertheless, we can use methods adapted for this purpose as shown in \cite{garcia2022jordan-I,garcia2022jordan-II,garcia2024jordan-III,Ahn_2021}.
    \item For constant Yang-Baxter solutions in higher dimensions, these methods go through with the SUSY algebras replaced by parasupersymmetry algebras. A possible place for testing this is in a recent work \cite{pourkia2018solutions} which claims to find solutions in higher dimensions as well.
    \item It would also be interesting to find next nearest-neighbor spin chains using the SUSY methods suitably modified to cover this setting. These can indeed be rewarding as they can be compared with very recent works on spin $\frac{1}{2}$ systems \cite{Shiraishi:2025psk}. For higher spins, see \cite{Shiraishi:2025tjn}.
 %   \item The methods used in this paper produce regular or the almost regular solutions as defined here. Non-regular solutions can also be obtained with appropriate modifications to the Baxterization methods used here. Such solutions can be compared with recent works on non-regular solutions \cite{garkun2024new,deLeeuw:2024zqj}.
    \item In recent times, machine learning methods have been utilized to extract solutions of the QYBE \cite{Lal:2023dkj,Lal:2025nmf}. These require the creation of test data sets in different-dimensional Hilbert spaces to develop good models. The higher spin solutions developed in this work can be used to create such data sets. 
\end{enumerate}

%%%%%%%%%%%%%%%%%%%%%%%%%%%%%%%%%%%%%%%%%%%%%%%%%%
%\section*{Data Availability}
%\label{sec:da}
%%%%%%%%%%%%%%%%%%%%%%%%%%%%%%%%%%%%%%%%%%%%%%%%%%
 %We declare that there is no data associated to this manuscript.

%%%%%%%%%%%%%%%%%%%%%%%%%%%%%%%%%%%%%%%%%%%%%%%%%%%%%%%%%%%%
%\section*{Conflict of Interest Statement}
%\label{sec:ci}
%%%%%%%%%%%%%%%%%%%%%%%%%%%%%%%%%%%%%%%%%%%%%%%%%%%%%%%%%%%%
%We, the Authors declare that, there are no financial, commercial, personal, or institutional interests that conflict with this work.

%%%%%%%%%%%%%%%%%%%%%%%%%%%%%%%%
\section*{Acknowledgments}
%\label{sec:}
%%%%%%%%%%%%%%%%%%%%%%%%%%%%%%%%
We thank the anonymous Referees of Journal of Physics A, for their valuable comments and suggestions to improve the manuscript.
VK is funded by the U.S. Department of Energy, Office of Science, National Quantum Information Science Research Centers, Co-Design Center for Quantum Advantage ($C^2QA$) under Contract No. DE-SC0012704.

\bibliographystyle{acm}
\normalem
\bibliography{refs}

\begin{thebibliography}{10}

\bibitem{Ahn_2021}
{\sc Ahn, C., and Staudacher, M.}
\newblock {The integrable (hyper)eclectic spin chain}.
\newblock {\em Journal of High Energy Physics 2021}, 2 (Feb. 2021).

\bibitem{Arnaudon_2003}
{\sc Arnaudon, D., Chakrabarti, A., Dobrev, V.~K., and Mihov, S.~G.}
\newblock {Spectral Decomposition and Baxterization of Exotic Bialgebras and Associated Noncommutative Geometries}.
\newblock {\em International Journal of Modern Physics A 18}, 23 (Sept. 2003), 4201–4213.

\bibitem{AuYang2016About3Y}
{\sc Au-Yang, H., and Perk, J. H.~H.}
\newblock About 30 years of integrable chiral potts model, quantum groups at roots of unity and cyclic hypergeometric functions.
\newblock {\em arXiv: Mathematical Physics\/} (2016).

\bibitem{axler2024linear}
{\sc Axler, S.}
\newblock {\em Linear algebra done right}.
\newblock Springer Nature, 2024.

\bibitem{bargheer2009long}
{\sc Bargheer, T., Beisert, N., and Loebbert, F.}
\newblock {Long-range deformations for integrable spin chains}.
\newblock {\em Journal of Physics A: Mathematical and Theoretical 42}, 28 (2009), 285205.

\bibitem{BAXTER1988138}
{\sc Baxter, R., Perk, J., and Au-Yang, H.}
\newblock New solutions of the star-triangle relations for the chiral potts model.
\newblock {\em Physics Letters A 128}, 3 (1988), 138--142.

\bibitem{BAXTER1972193}
{\sc Baxter, R.~J.}
\newblock {Partition function of the Eight-Vertex lattice model}.
\newblock {\em Annals of Physics 70}, 1 (1972), 193--228.

\bibitem{Caux_2011}
{\sc Caux, J.-S., and Mossel, J.}
\newblock Remarks on the notion of quantum integrability.
\newblock {\em Journal of Statistical Mechanics: Theory and Experiment 2011}, 02 (Feb. 2011), P02023.

\bibitem{Cooper_1995}
{\sc Cooper, F., Khare, A., and Sukhatme, U.}
\newblock {Supersymmetry and quantum mechanics}.
\newblock {\em Physics Reports 251}, 5–6 (Jan. 1995), 267–385.

\bibitem{deLeeuw2024}
{\sc Corcoran, L., and de~Leeuw, M.}
\newblock {All regular $4 \times 4$ solutions of the Yang-Baxter equation}.
\newblock {\em SciPost Phys. Core 7\/} (2024), 045.

\bibitem{Crampe_2016}
{\sc Crampe, N., Frappat, L., Ragoucy, E., and Vanicat, M.}
\newblock {A New Braid-like Algebra for Baxterisation}.
\newblock {\em Communications in Mathematical Physics 349}, 1 (Oct. 2016), 271–283.

\bibitem{crampe_2019}
{\sc Crampe, N., Ragoucy, E., and Vanicat, M.}
\newblock {Back to Baxterisation}.
\newblock {\em Communications in Mathematical Physics 365}, 3 (2019), 1079--1090.

\bibitem{de2019classifying}
{\sc De~Leeuw, M., Pribytok, A., and Ryan, P.}
\newblock {Classifying integrable spin-1/2 chains with nearest neighbour interactions}.
\newblock {\em Journal of Physics A: Mathematical and Theoretical 52}, 50 (2019), 505201.

\bibitem{Essler_Frahm_Gohmann_Klümper_Korepin_2005}
{\sc Essler, F. H.~L., Frahm, H., G\"{o}hmann, F., Klümper, A., and Korepin, V.~E.}
\newblock {\em The One-Dimensional Hubbard Model}.
\newblock Cambridge University Press, 2005.

\bibitem{Fei:1991zn}
{\sc Fei, S.-M., Guo, H.-Y., and Shi, H.}
\newblock {Multiparameter solutions of the Yang-Baxter equation}.
\newblock {\em J. Phys. A 25\/} (1992), 2711--2720.

\bibitem{garcia2022jordan-II}
{\sc Garc{\'\i}a, J. M.~N.}
\newblock {Jordan blocks and the Bethe Ansatz II: The eclectic spin chain beyond K= 1}.
\newblock {\em Journal of High Energy Physics 2022}, 12 (2022), 1--30.

\bibitem{garcia2024jordan-III}
{\sc Garc{\'\i}a, J. M.~N.}
\newblock {Jordan blocks and the Bethe ansatz III: Class 5 model and its symmetries}.
\newblock {\em Nuclear Physics B 998\/} (2024), 116419.

\bibitem{garcia2022jordan-I}
{\sc Garc{\'\i}a, J. M.~N., and Wyss, L.}
\newblock {Jordan blocks and the Bethe Ansatz I: The eclectic spin chain as a limit}.
\newblock {\em Nuclear Physics B 981\/} (2022), 115860.

\bibitem{ge-baxterization}
{\sc Ge, M.-L., Xue, K., and Wu, Y.-S.}
\newblock {Yang-Baxterization and algebraic structures}.
\newblock {\em Braid Group, Knot Theory and Statistical Mechanics II\/}, 130--152.

\bibitem{halverson2004partitionalgebras}
{\sc Halverson, T., and Ram, A.}
\newblock {Partition Algebras}.
\newblock {\em arXiv:math/0401314\/} (2004).

\bibitem{HIETARINTA-PLA}
{\sc Hietarinta, J.}
\newblock {All solutions to the constant quantum Yang-Baxter equation in two dimensions}.
\newblock {\em Physics Letters A 165}, 3 (1992), 245--251.

\bibitem{hietarinta1993-JMP-Long}
{\sc Hietarinta, J.}
\newblock {Solving the two‐dimensional constant quantum Yang–Baxter equation}.
\newblock {\em Journal of Mathematical Physics 34}, 5 (05 1993), 1725--1756.

\bibitem{Hlavaty_1987}
{\sc Hlavaty, L.}
\newblock {Unusual solutions to the Yang-Baxter equation}.
\newblock {\em Journal of Physics A: Mathematical and General 20}, 7 (may 1987), 1661.

\bibitem{horn2012matrix}
{\sc Horn, R.~A., and Johnson, C.~R.}
\newblock {\em Matrix analysis}.
\newblock Cambridge university press, 2012.

\bibitem{Jones1990}
{\sc Jones, V. F.~R.}
\newblock {\em {Baxterization}}.
\newblock Springer US, Boston, MA, 1990, pp.~5--11.

\bibitem{Korepin1993QuantumIS}
{\sc Korepin, V.~E., Bogoliubov, N.~M., and Izergin, A.~G.}
\newblock {Quantum Inverse Scattering Method and Correlation Functions}.
\newblock Cambridge university press.

\bibitem{krichever1981baxter}
{\sc Krichever, I.~M.}
\newblock {Baxter's equations and algebraic geometry}.
\newblock {\em Functional Analysis and Its Applications 15}, 2 (1981), 92--103.

\bibitem{Kruczenski2004}
{\sc Kruczenski, M.}
\newblock {Spin Chains and String Theory}.
\newblock {\em Phys. Rev. Lett. 93\/} (Oct 2004), 161602.

\bibitem{kulish-baxterization}
{\sc Kulish, P.~P., Reshetikhin, N.~Y., and Sklyanin, E.~K.}
\newblock {Yang-Baxter equation and representation theory: I}.
\newblock {\em Letters in Mathematical Physics 5}, 5 (1981), 393--403.

\bibitem{Lal:2023dkj}
{\sc Lal, S., Majumder, S., and Sobko, E.}
\newblock {The R-mAtrIx Net}.
\newblock {\em Mach. Learn. Sci. Tech. 5}, 3 (2024), 035003.

\bibitem{Lal:2025nmf}
{\sc Lal, S., Majumder, S., and Sobko, E.}
\newblock {Deep Learning based discovery of Integrable Systems}.
\newblock {\em arXiv:2503.10469\/} (3 2025).

\bibitem{li1993yang}
{\sc Li, Y.-Q.}
\newblock {Yang Baxterization}.
\newblock {\em Journal of mathematical physics 34}, 2 (1993), 757.

\bibitem{loebbert2016lectures}
{\sc Loebbert, F.}
\newblock {Lectures on Yangian symmetry}.
\newblock {\em Journal of Physics A: Mathematical and Theoretical 49}, 32 (2016), 323002.

\bibitem{maity2025non}
{\sc Maity, S., Padmanabhan, P., and Korepin, V.}
\newblock {Non-hermitian integrable systems from constant non-invertible solutions of the Yang-Baxter equation}.
\newblock {\em Journal of High Energy Physics 2025}, 5 (2025), 1--31.

\bibitem{MSPK-Hietarinta}
{\sc Maity, S., Singh, V.~K., Padmanabhan, P., and Korepin, V.}
\newblock {Algebraic classification of Hietarinta's solutions of Yang-Baxter equations: invertible 4 ×4 operators}.
\newblock {\em Journal of High Energy Physics 2024}, 12 (2024), 67.

\bibitem{martin1994temperley}
{\sc Martin, P.}
\newblock {Temperley-Lieb algebras for non-planar statistical mechanics—the partition algebra construction}.
\newblock {\em Journal of Knot Theory and its Ramifications 3}, 01 (1994), 51--82.

\bibitem{minahan2012review}
{\sc Minahan, J.~A.}
\newblock {Review of AdS/CFT Integrability, Chapter I. 1: Spin Chains in Super Yang-Mills}.
\newblock {\em Letters in Mathematical Physics 99}, 1 (2012), 33--58.

\bibitem{padmanabhan2024solving}
{\sc Padmanabhan, P., and Korepin, V.}
\newblock {Solving the Yang-Baxter, tetrahedron and higher simplex equations using Clifford algebras}.
\newblock {\em Nuclear Physics B 1007\/} (Oct. 2024), 116664.

\bibitem{Padmanabhan_2020}
{\sc Padmanabhan, P., Sugino, F., and Trancanelli, D.}
\newblock {Braiding quantum gates from partition algebras}.
\newblock {\em Quantum 4\/} (Aug. 2020), 311.

\bibitem{pourkia2018solutions}
{\sc Pourkia, A.}
\newblock {Solutions to the constant Yang-Baxter equation in all dimensions}.
\newblock {\em arXiv:1806.08400\/} (2018).

\bibitem{Pozsgay:2021rwc}
{\sc Pozsgay, B., Gombor, T., and Hutsalyuk, A.}
\newblock {Integrable hard-rod deformation of the Heisenberg spin chains}.
\newblock {\em Phys. Rev. E 104}, 6 (2021), 064124.

\bibitem{Shiraishi:2025psk}
{\sc Shiraishi, N.}
\newblock {Complete classification of integrability and non-integrability of S=1/2 spin chains with symmetric next-nearest-neighbor interaction}.
\newblock {\em arXiv:2501.15506\/} (1 2025).

\bibitem{Shiraishi:2025tjn}
{\sc Shiraishi, N., and Yamaguchi, M.}
\newblock {Dichotomy theorem distinguishing non-integrability and the lowest-order Yang-Baxter equation for isotropic spin chains}.
\newblock {\em arXiv:2504.14315\/} (4 2025).

\bibitem{Sinha_2025}
{\sc Sinha, A., Justin, T., Padmanabhan, P., and Korepin, V.}
\newblock The yang–baxter integrability of the critical ising chain.
\newblock {\em Journal of Statistical Mechanics: Theory and Experiment 2025}, 10 (Oct. 2025), 103102.

\bibitem{takhtadzhyan1979}
{\sc Sklyanin, E.~K., Takhtadzhyan, L.~A., and Faddeev, L.~D.}
\newblock {Quantum inverse problem method. I}.
\newblock {\em Theoretical and Mathematical Physics 40}, 2 (1979), 688--706.

\bibitem{slavnov2019algebraicbetheansatz}
{\sc Slavnov, N.~A.}
\newblock {Algebraic Bethe ansatz}.
\newblock {\em arXiv:1804.07350 [math-ph]\/} (2019).

\bibitem{Sogo}
{\sc Sogo, K., Uchinami, M., Akutsu, Y., and Wadati, M.}
\newblock {Classification of Exactly Solvable Two-Component Models: }.
\newblock {\em Progress of Theoretical Physics 68}, 2 (08 1982), 508--526.

\bibitem{staudacher2012review}
{\sc Staudacher, M.}
\newblock {Review of AdS/CFT integrability, chapter III. 1: Bethe ans{\"a}tze and the R-matrix formalism}.
\newblock {\em Letters in Mathematical Physics 99}, 1 (2012), 191--208.

\bibitem{Takhtadzhan_1979}
{\sc Takhtadzhan, L.~A., and Faddeev, L.~D.}
\newblock {The Quantum Method of the Inverse Problem and the Heisenberg XYZ Model}.
\newblock {\em Russian Mathematical Surveys 34}, 5 (oct 1979), 11.

\bibitem{Vanicat2017IntegrableTL}
{\sc Vanicat, M., Zadnik, L., and Prosen, T.}
\newblock {Integrable Trotterization: Local Conservation Laws and Boundary Driving.}
\newblock {\em Physical review letters 121 3\/} (2017), 030606.

\bibitem{vieira2018solving}
{\sc Vieira, R.~S.}
\newblock {Solving and classifying the solutions of the Yang-Baxter equation through a differential approach. Two-state systems}.
\newblock {\em Journal of High Energy Physics 2018}, 10 (2018), 1--50.

\bibitem{Witten1981DynamicalBO}
{\sc Witten, E.}
\newblock {Dynamical Breaking of Supersymmetry}.
\newblock {\em Nuclear Physics 188\/} (1981), 513--554.

\bibitem{YangCN1967}
{\sc Yang, C.~N.}
\newblock {Some Exact Results for the Many-Body Problem in one Dimension with Repulsive Delta-Function Interaction}.
\newblock {\em Phys. Rev. Lett. 19\/} (Dec 1967), 1312--1315.

\bibitem{ZHANG1991625}
{\sc Zhang, R., Gould, M., and Bracken, A.}
\newblock {From representations of the braid group to solutions of the Yang-Baxter equation}.
\newblock {\em Nuclear Physics B 354}, 2 (1991), 625--652.

\bibitem{Zhang_1991}
{\sc Zhang, R.~B.}
\newblock {Multiparameter dependent solutions of the Yang-Baxter equation}.
\newblock {\em Journal of Physics A: Mathematical and General 24}, 10 (may 1991), L535.

\end{thebibliography}

\end{document}